\documentclass[]{aa}  
\usepackage{natbib}
\bibpunct{(}{)}{;}{a}{}{,} 
\usepackage{graphicx}
\usepackage{txfonts}
\usepackage{color}

\begin{document} 

\title{First scattered light detection of a nearly edge-on transition disk around the T Tauri star RY Lup}

\author{M. Langlois\inst{1,2} A. Pohl\inst{3,4}, A.-M. Lagrange\inst{5} , A.- L. Maire\inst{3}, D. Mesa\inst{6}, A. Boccaletti\inst{7}, R. Gratton\inst{6}, L. Denneulin\inst{1,2}, H. Klahr\inst{3}, A. Vigan\inst{2}, M. Benisty\inst{5}, C. Dominik\inst{8},  M. Bonnefoy\inst{5}, F. Menard\inst{5}, H. Avenhaus\inst{9}, A. Cheetham\inst{10}, R. Van Boekel\inst{3}, J. de Boer\inst{11}, G. Chauvin\inst{5}, S. Desidera\inst{6} , M. Feldt\inst{3}, R. Galicher\inst{7}, C. Ginski\inst{11}, J. Girard\inst{2}, T. Henning\inst{3}, M. Janson\inst{12,3}, T. Kopytova\inst{7}, Q. Kral\inst{7,15},  R. Ligi\inst{2}, S. Messina\inst{13}, S. Peretti\inst{10}, C. Pinte\inst{5}, E. Sissa\inst{6},  T Stolker\inst{14}, A. Zurlo\inst{2}, Y. Magnard\inst{5}, P. Blanchard\inst{2}, T. Buey\inst{7}, M. Suarez\inst{16}, E. Cascone\inst{17}, O. Moller-Nilsson\inst{3}, L. Weber\inst{10}, C. Petit\inst{18}, J. Pragt\inst{19}}

\institute{CRAL, UMR 5574, CNRS, Universit\'{e} Lyon 1, 9 avenue Charles Andr\'{e}, 69561 Saint Genis Laval Cedex, France\\
        \email{maud.langlois@univ-lyon1.fr@univ-lyon1.fr}
\and    
        Aix Marseille Universit\'{e}, CNRS, LAM (Laboratoire d'Astrophysique de Marseille) UMR 7326, 13388, Marseille, France
\and
Max Planck Institute for Astronomy, K\"onigstuhl 17, D-69117 Heidelberg, Germany
\and
        Heidelberg University, Institute of Theoretical Astrophysics, Albert-Ueberle-Str. 2, D-69120 Heidelberg, Germany
\and
        Univ. Grenoble Alpes, CNRS, IPAG, F-38000 Grenoble, France
\and
    INAF-Osservatorio Astronomico di Padova, Vicolo dell Osservatorio 5, 35122 Padova, Italy
    \and
 LESIA, Observatoire de Paris, PSL Research University, CNRS, Sorbonne Universit\'{e}s, UPMC Univ. Paris 06, Univ. Paris Diderot, Sorbonne Paris Cit\'{e}, 5 place Jules Janssen, 92195 Meudon, France
 \and
Anton Pannekoek Institute for Astronomy, University of Amsterdam, Science Park 904, 1098 XH Amsterdam, The Netherland
 \and
 ETH Zurich, Institute for Astronomy, Wolfgang Pauli Strasse 27, CH 8093, Zurich, Switzerland
 \and
        Observatoire Astronomique de l'Universit\'{e} de Gen\`{e}ve, 51 Ch. des Maillettes, 1290 Versoix, Switzerland
 \and
        Leiden Observatory, Leiden University, P.O. Box 9513, 2300 RA Leiden, The Netherlands
 \and
        Department of Astronomy, Stockholm University, AlbaNova University Center, 10691 Stockholm, Sweden
\and
        Osservatorio Astrofisico di Catania, Via S.Sofia 78, 95123 Catania ITALY
\and
        Anton Pannekoek Institute for Astronomy, University of Amsterdam, Science Park 904, 1098 XH Amsterdam, The Netherlands
\and
        Institute of Astronomy, University of Cambridge, Madingley Road, Cambridge CB3 0HA, UK
\and
        European Southern Observatory, Karl-Schwarzschild-Strasse 2, 85748 Garching bei Munchen, Germany
\and
        INAF, Astrophysical Observatory of Capodimonte, Salita Moiariello 16, 80131 Napoli, Italy
\and
        ONERA, 29 avenue de la Division Leclerc, 92322 Chatillon Cedex, France
\and
        NOVA Optical-Infrared Instrumentation Group at ASTRON, Oude Hoogeveensedijk 4, 7991 PD Dwingeloo, The Netherlands
}

\abstract
   {Transition disks are considered sites of ongoing planet formation, and their dust and gas distributions could be signposts of embedded planets. The transition disk around the T Tauri star RY Lup has an inner dust cavity and displays a strong silicate emission feature.}
   {Using high-resolution imaging we  study the disk geometry, including non-axisymmetric features, and its surface dust grain, to gain a better understanding of the disk evolutionary process. Moreover, we   search for companion candidates, possibly connected to the disk.}
   {We obtained high-contrast and high angular resolution data in the near-infrared with the VLT/SPHERE extreme adaptive optics instrument whose goal is  to study the planet formation by detecting and characterizing these planets and their formation environments through direct imaging. We performed polarimetric imaging of the RY~Lup disk with IRDIS (at 1.6 $\muup$m), and obtained intensity images with the IRDIS dual-band imaging camera simultaneously with the IFS spectro-imager (0.9--1.3 $\muup$m).}
   {We resolved for the first time the scattered light from the nearly edge-on  circumstellar disk around RY~Lup, at projected separations in the 100 \,au range. The shape of the disk and its sharp features are clearly detectable at wavelengths ranging from 0.9 to 1.6 $\muup$m. We show that the observed morphology can be interpreted as spiral arms in the disk. This interpretation is supported by in-depth numerical simulations. We also demonstrate that these features can be produced by one planet interacting with the disk. We also detect several point sources which are classified as probable background objects.}
  {}
   
 \titlerunning{Transition Disk around the T Tauri star RY~Lup}
 \authorrunning{Langlois et al.}
%
%
\maketitle
%
%

\section{Introduction}     \label{sec:introduction}
High-resolution and high-contrast imaging capabilities provided by the new generation of adaptive optics based instruments such as the Spectro-Polarimeter High-contrast Exoplanet REsearch (SPHERE) instrument \citep{beuzit2008} and the Gemini Planer Imager (GPI)  \citep{macintosh2014} open the path to directly image new protoplanetary disks in scattered light. Protoplanetary disks are optically thick in the optical and near-infrared (NIR), so that scattered light imaging probes micron-sized dust grains in the disk's surface layer, while mm observations trace larger, mm-sized grains located close to the disk's mid-plane. Transition disks (TD) with gas and dust gaps  \citep[e.g.][]{brown2007,merin2010,vanderMarel2016,ansdell2016,boekel2017} are particularly interesting since they might correspond to the stage where planet forming processes are active, thus we expect to see signposts of planet--disk interactions. The Atacama Large Millimeter Array (ALMA) now provides sufficient sensitivity and resolution at sub-mm wavelengths to detect these gaps \citep{fedele2017} and to estimate the mass of protoplanetary disks  around YSOs \citep[]{ansdell2016,trapman2017}, although the bulk mass (H$_2$ and He) is not directly observable. In recent observational studies of transition disks, giant gaps and cavities have also been directly imaged in scattered light \citep[e.g.][]{thalmann2010,thalmann2015, hashimoto2012, avenhaus2014,follette2015,ohta2016,stolker2016,benisty2017,pohl2017,boekel2017}. In addition, the detection of non-axisymmetric disk features that could be related to the presence of planets is fundamental in order to improve our current understanding of the disk evolution and the planet formation process. However, the small angular separations involved pose great observational challenges.
 
RY~Lup has been classified as a T Tauri G-type star showing type III variability located in the Upper Centaurus Lupus (UCL) association with an estimated age of 10 -- 20 Myr \citep[e.g.][]{manset2009,pecaut2012}.  However, a colder effective temperature of 5000 K and a best fit spectral type of K2  have been estimated by  \citet{biazzo2017} in the most recent literature based on X-Shooter spectroscopy. Its distance has been estimated by recent measurements using GAIA \citep[first GAIA data release,][]{gaia}, which  provided a distance estimate of 151pc $\pm$ 1 pc. RY~Lup is a TD, as confirmed by \cite{ansdell2016} using the ALMA data, but was not previously identified as such from its SED \citep{manset2009}. The spectral energy distribution of RY~Lup shows an infrared excess, as well as a modest UV excess \citep{evans1982,gahm1989}. Its strong 10 $\muup$m silicate emission feature, seen in its IRS spectrum \citep{kessler-silacci2006}, washes out the mid-IR dip in its broad-band fluxes.  
 
The photometric variability of the star has been revealed by Woods using Harvard plates in 1921. The star was classified as a typical RW Aur star (a variable eruptive T Tauri-type star) on the basis of its light curve. From a long-term study of the optical photometry of the star, \cite{gahm1989} reported significant photometric variations (up to 1.5 mag at visible wavelengths) on timescales of hours, a constant variability with a 3.75 day period over 50 years, and a decline in the mean and maximum photometric magnitudes over decades. This photometric behaviour of RY~Lup is typical of type III variability defined by variable stellar obscuration by circumstellar dust. These stars are characterized by an increase in the degree of linear polarization as the star becomes faint, supporting the idea that variable obscuration is responsible for the fading of the star. Historically the star has been observed to be redder when it is at a fainter magnitude \citep{evans1982,liseau1987,hutchinson1989,covino1992}, and the amplitude of the variations also decreases with wavelength. The intrinsic position angle of T Tauri star polarization is generally a function of both wavelength and time \citep{bastien1981,bastien1985}. These authors also noticed remarkably large and rapid variations in both polarization and position angle in RY~Lup. More recently, \cite{manset2009} has investigated  these photometric and polarimetric variabilities using simultaneous BV polarimetric and UBV photometric observations. They showed that the polarization is high (3.0\%) when the star is faint and red (V = 12.0 mag, B - V = 1.3 mag), and  low (0.5\%) when it is bright and bluer (V = 11.0 mag, B -V = 1.1 mag). The photometric and polarimetric variations share a common period of 3.75 d. They concluded that linear polarization is produced by dust scattering in an asymmetric (flat) circumstellar envelope, and they explained the photometric and polarimetric variations by an almost edge-on circumstellar disk that is warped close to the star, where it interacts with the star magnetosphere. They claim that the inhomogeneous disk matter contained in the warp could be corotating with the star and could partially occult it during part of the rotation period, which explains the dips in luminosity and the accompanying increase in polarization. However, the most recent estimate of $v_{sini}=16.3 \pm 5.3 km/s$ from \citep{alcala2017} is smaller than earlier measured. This implies that the stellar rotation axis has an inclination likely < 70 deg. Therefore, the disk is probably misaligned with respect to the star's equatorial plane.

Recent ALMA high-resolution sub-mm observations of RY~Lup in the 890 micron dust continuum \citep{ansdell2016} provide a detailed map of the spatial distribution of large, mm-sized dust, at a linear spatial resolution of 35 \,au. They show a clear signature of an inner mm dust cavity with a diameter of 0.8'' (~60 \,au) and a clearly resolved dust ring. These continuum emission measurements constrain M$_{dust}$ to less than 0.2 M$_{\odot}$ and additional CO isotopologue emission ($_{13}$CO and C$_{18}$O 3-2 lines) constrain M$_{gas}$ to less than 2.6 M$_{\odot}$.The gas-to-dust ratio has been estimated to be 5--50 when using their parameterized model framework. Although the gas mass estimation may be underestimated due to their assumption of an ISM-like abundance, the weak CO isotopologue emission indicates rapid disk evolution, either directly in the gas-to-dust ratio or chemically via permanent loss of volatiles to solids. All previous works on RY~Lup are consistent with a system comprising a star surrounded by a nearly edge-on disk. \cite{manset2009} estimate the mass of RY~Lup to M$_{\star}$/M$_{\odot}$=1.4, with an age of 12 Myr. This region's proximity and age makes it ideal for a baseline study of disk properties.

 \cite{Dodson2011} demonstrate that  transitional disks with wide cavities  (>15 AU) and high accretion rate greater than ${10^{-8}}  M_{\odot}/Yr$ presents unambiguous evidence for tidal clearing by multiple  planets. The cavity formation can only occur once evaporation rates exceed accretion rates, as otherwise the hole will be replenished by accretion. Photoevaporation models do not appear to be able to explain the existence of disks with such characteristics. The RY Lup disk clearly falls in that category with an estimated mass accretion rate ${10^{-8.2}}  M_{\odot}/Yr$ from \cite{gahm1993}.

As part of the SPHERE guaranteed time observations dedicated to a large survey to search for planets and to study circumstellar disks around members of young and nearby associations, we recently recorded high-contrast images of RY~Lup. The data resolve the disk for the first time in intensity in the near-infrared and in polarized light. This article  presents the observational results, and aims to develop qualitative arguments on the detected disk. In parallel to the observation, detailed numerical hydrodynamical simulations in the context of planet--disk interaction in combination with radiative transfer calculations were performed to study the disk morphology. We first describe the observations and the data analysis (Sect.~\ref{sec:observations}), then the results obtained on the disk and the search for planets in this system (Sect.~\ref{sec:Results}). This is followed by detailed numerical modelling of the disk (Sect.~\ref{sec:diskmod}).

\section{Observations and data reduction}    \label{sec:observations}

 \begin{table*}[t]
\caption{Observational parameters}              
\label{table:data}      
\centering                                      
\begin{tabular}{c c c c c c c}          
\hline\hline                        
Date & Mode & DIT(s) x NDIT  & Seeing ('')  & Strehl (\%) & ADI FOV rotation (deg) \\    
\hline                                   
 2016-04-16 & IRDIS DB-H23 & 64 x 80    & 0.25 &  78 & 71\\
\hline  
2016-04-16 & IFS YJ &  64 x 80  & 0.25 & 78 & 74  \\
\hline
 2016-05-27   & IRDIS DPI BB-H& 32 x 256 & 0.3 & 68   \\ 
\hline      
\end{tabular}
\end{table*}

\subsection{Observations}

All observations were part of the SPHERE consortium guaranteed time programme (SPHERE High-Contrast Imaging Survey for Exoplanets and SPHERE Disk Survey). The Infra-Red Dual-beam Imager and Spectrograph \citep[IRDIS,][]{Dohlen2008} and Integral Field Spectrograph \citep[IFS,][]{claudi2008} were used. The first RY~Lup datasets  were recorded on  16  April 2016 with the IRDIFS instrumental configuration, where observations are simultaneously carried out with the IRDIS subsystem dual-band imaging (DBI) mode using H2,H3 filters respectively centred  at 1.59 and 1.67 $\muup$m \citep{vigan2010}, and the Integral Field Spectrometer operating in YJ bands (0.95 -- 1.35 $\muup$m) with a spectral resolution of 42 in Y band to 40 in J Band. Following the discovery of the disk in intensity, we obtained Dual-band Polarimetric Imaging \citep[DPI,][]{langlois2010} observations in field stabilized mode on  27 May 2016. 
IRDIS provides a 10--12 $\arcsec$ square field of view (1500--1800 square au, given the star's distance 151 pc, using 12.25 mas/pixel platescale i.e.  1.85 au per pixel). The IFS dataset consists of 21000 spectra each spread over 5.1 x 41 pixels on the detector. After extraction, the FoV is 1.7$\arcsec$ square and the spaxel size is 7.46 x 7.46 mas, i.e. 1.13 au per pixel. We used an apodized Lyot coronagraph \citep{soummer2005,boccaletti2008,carbillet2011,guerri2011}, including a 185 mas focal mask, as well as a pupil mask (N-ALC-YJH-S). The IRDIFS observations were performed in pupil stabilized mode in order to perform Angular Differential Imaging (ADI) post-processing \citep{marois2006}. The field rotation for this set of data is 73 degrees. The observing sequences for both IRDIS and IFS can be summarized as follows: 1) point spread function (PSF) imaging, including a small star offset to displace the star away from the coronagraphic mask in order to record the unsaturated PSF which provides relative photometric calibration (we note that a neutral density filter was also inserted to avoid saturation); 2) imaging of the star behind the coronagraphic mask, with four crosswise faint replicas of the star artificially generated using the deformable mirror \citep{langlois2013} used for fine monitoring of the star centre; and 3) science coronographic sequence. At the end of the observing sequence, we
repeated the first and second steps followed by a calibration of the sky background, with detector integration times (DITs) corresponding to the DITs of the coronagraphic observations. Finally, the true north (TN) and pixel plate scales were measured using the astrometric calibrator NGC3603 \citep{Khorrami2017} observed as part of the standard SPHERE  calibration routines performed for the Guaranteed Time Observations (GTO) survey \citep{maire2016}. More details on the datasets are presented in Table \ref{table:data}. The observing conditions were very good for all observing sequences. The Strehl ratio estimation provided in Table \ref{table:data} is based on an extrapolation of the phase variance deduced from the reconstruction of the SPHERE Adaptive Optics system open-loop data using deformable mirror and tip-tilt voltages, and wavefront sensor closed-loop data \citep{fusco2004}. The observing conditions and the different data reduction methods for each dataset are described in more detail in Sects.  \ref{sec:IRDIFS obs} and \ref{sec:DPI obs}.

\subsection{Intensity images using IRDIS-DBI (H2,H3 bands) and IFS (YJ band)} \label{sec:IRDIFS obs}
\begin{figure*}[hbtp]
  \centering
  \includegraphics[width=80mm]{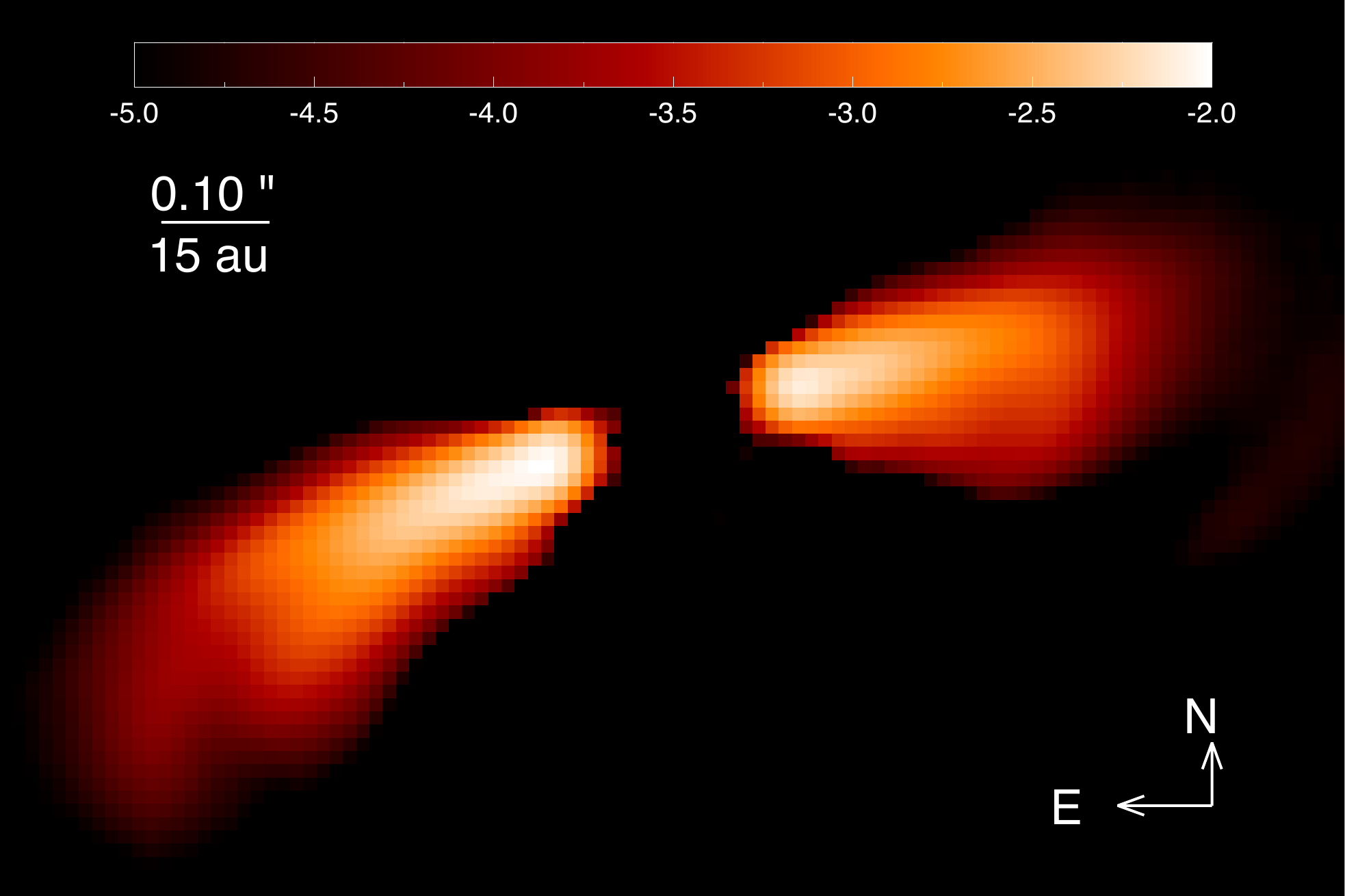}
  \includegraphics[width=80mm]{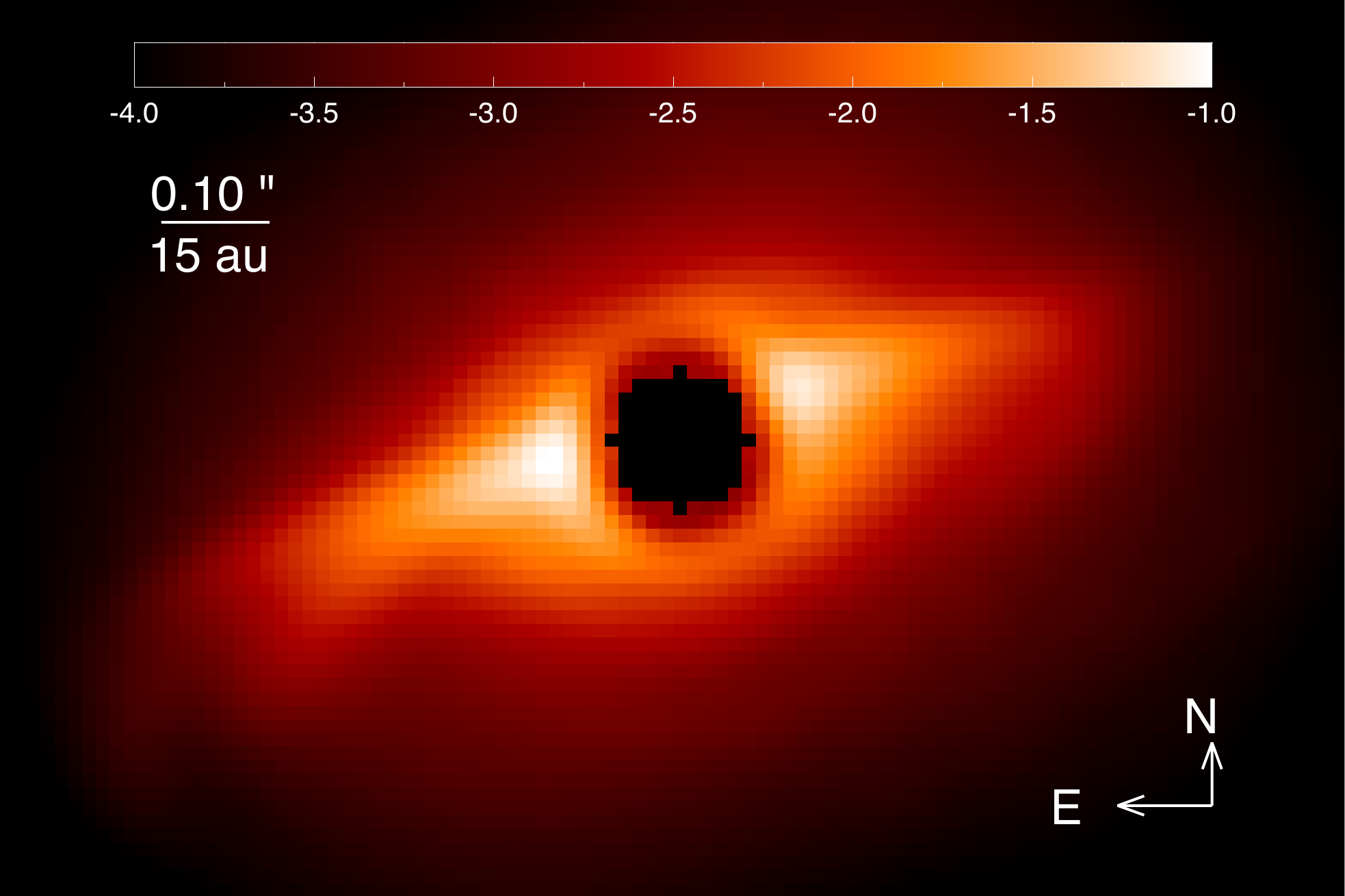}
   \includegraphics[width=80mm]{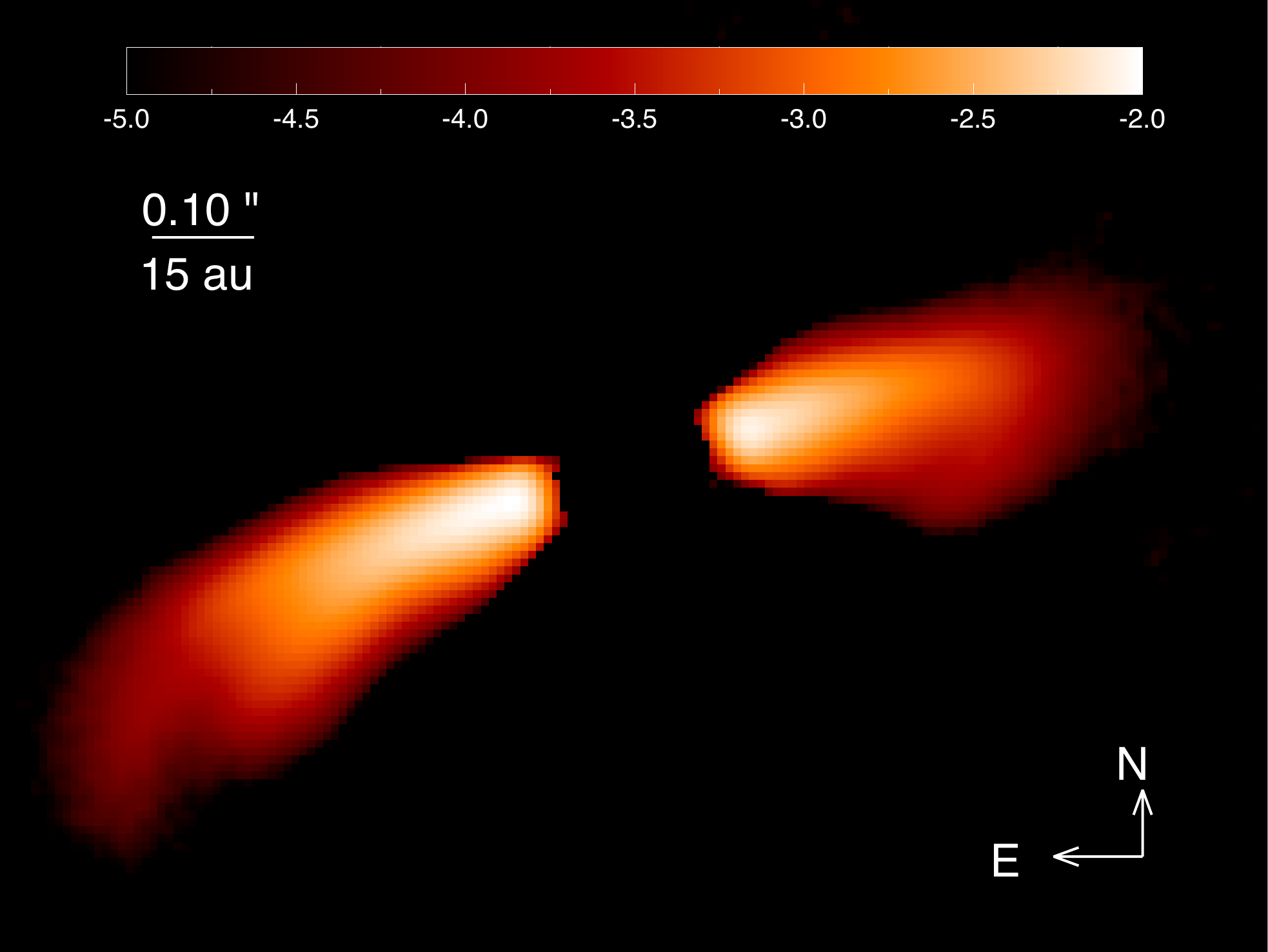}
   \includegraphics[width=80mm]{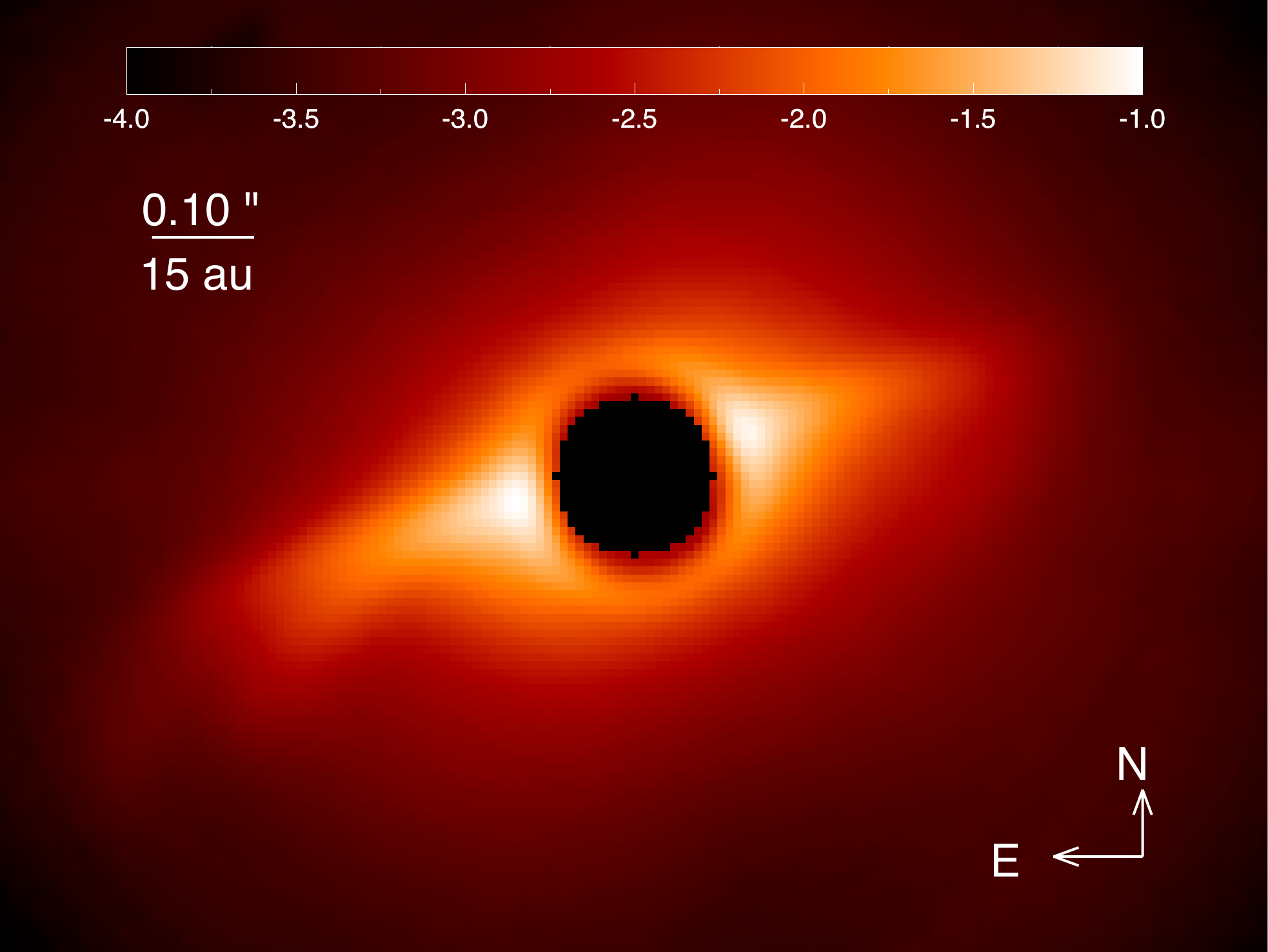}
  \caption[]{\label{fig:CADI_IRD}
  IRDIS cADI (top left) and noADI (top right) H2+H3 image normalized to the maximum of the unsaturated non-coronagraphic PSF showing the close-in environment of RY~Lup. cADI (bottom left)  and noADI (bottom right) IFS YJ image normalized to the maximum of the unsaturated non-coronagraphic PSF showing the close-in environment of RY~Lup. The dark central region corresponds to the area masked by the coronagraph.  The intensity scale is logarithmic. The orientation is standard:  north is up and the east is towards the left}
\end{figure*}

The IRDIS  data were first corrected for cosmetics and sky background using the SPHERE Data Reduction and Handling (DRH) pipeline \citep{pavlov2008} implemented at the SPHERE Data Centre. The outputs include cubes of left and right images recentred onto a common origin (the centre of the field rotation) using the satellite spots. This preprocessing also includes background subtraction, bad-pixel interpolation, flat-field correction, distortion correction, and wavelength calibration (for IFS). After these first steps, the best frames were selected according to their quality, leading to the use of 72 frames out of 80 for IRDIS and of all the available frames for IFS. The IRDIS frame selection criteria is based on the central coronagraphic spot flux  being within a $\pm$2 $\sigma$ deviation from the median value of the full sequence. 

\begin{figure*}[hbtp]
  \centering
  \includegraphics[width=80mm]{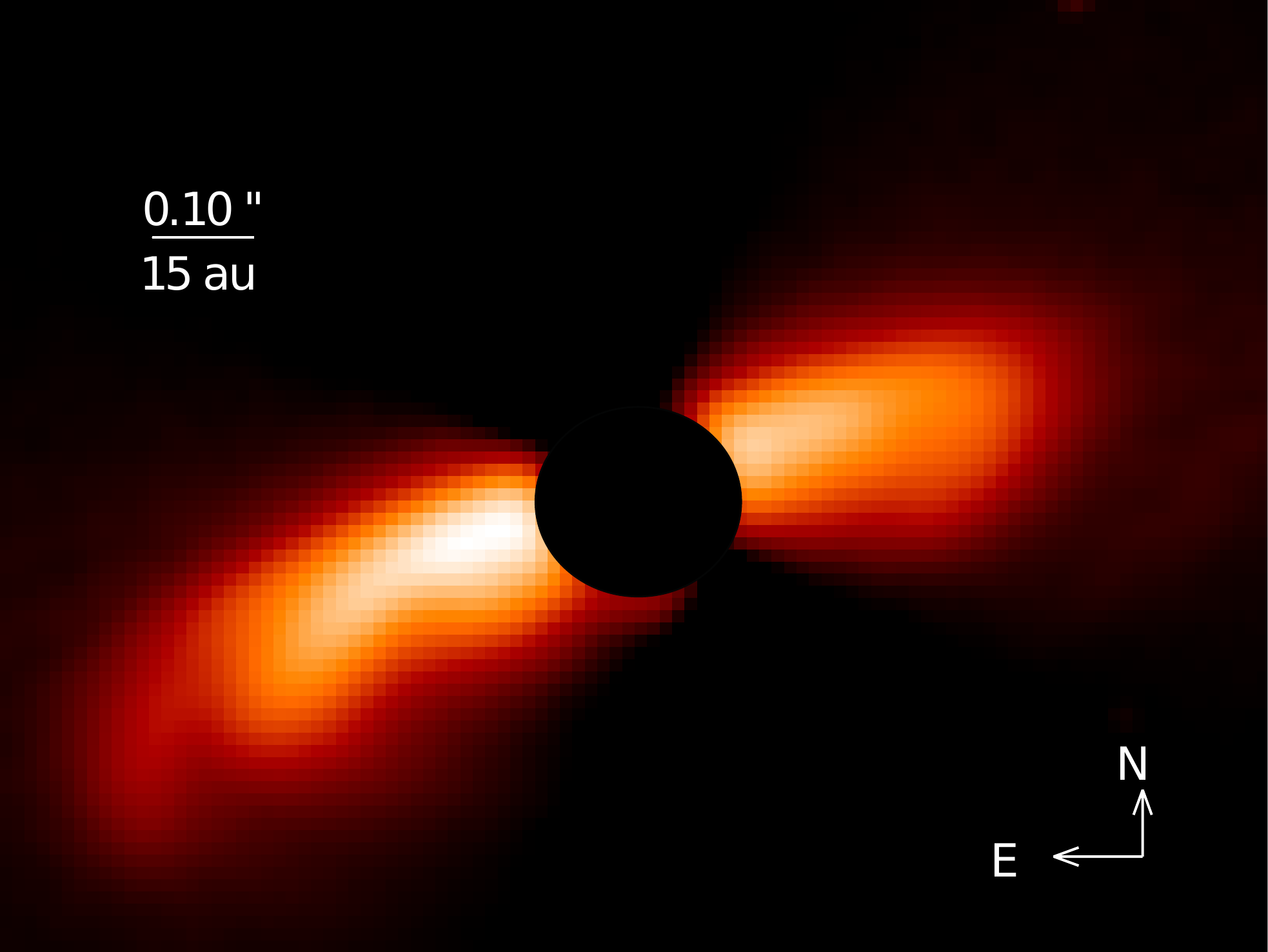}
  \includegraphics[width=80mm]{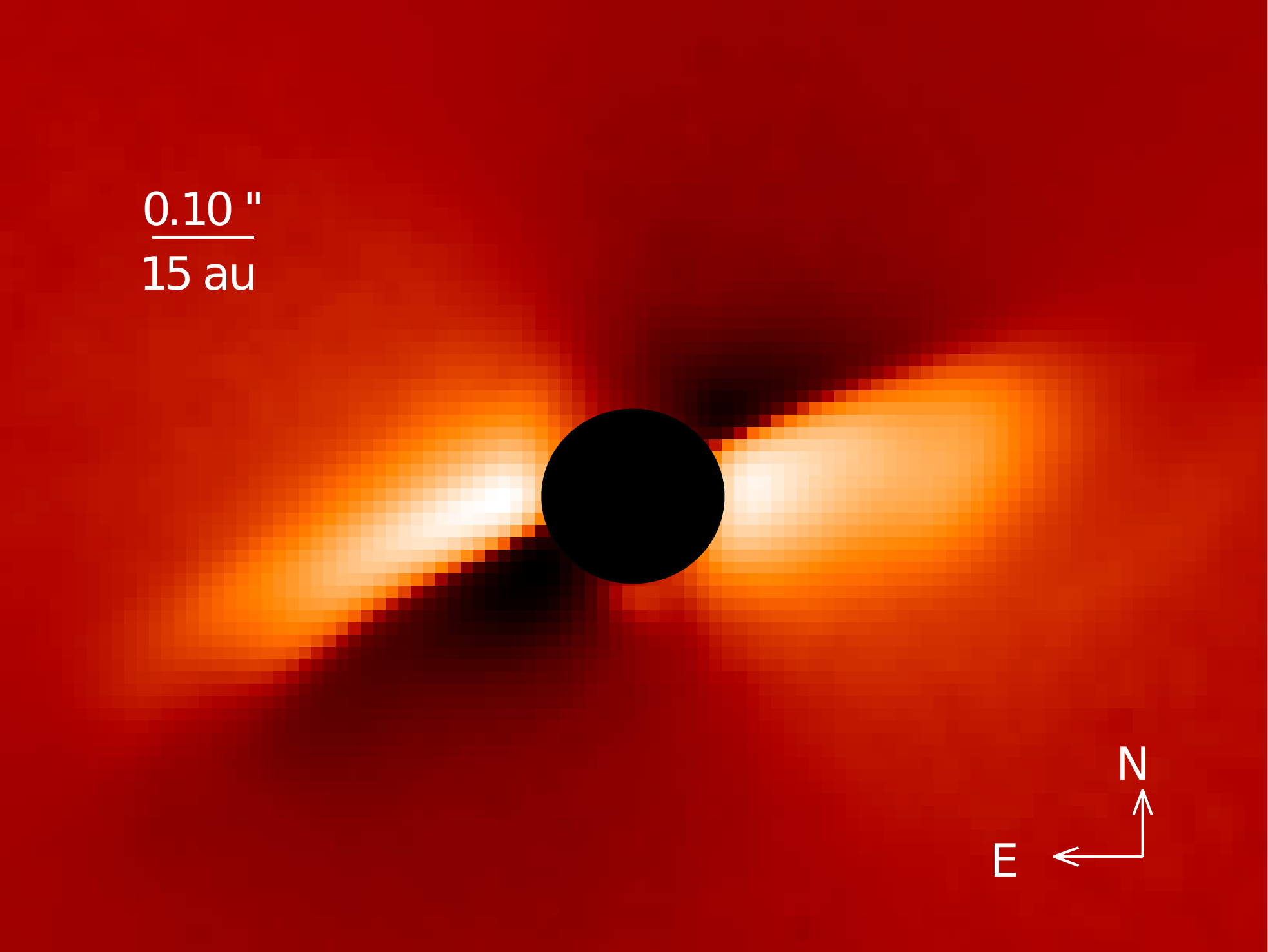}
  \caption[]{\label{fig:pola_image}
    IRDIS DPI Q$_\varphi$ (left) and U$_\varphi$ (right) images showing the close environment of RY~Lup. The dark central region corresponds to the area masked by the coronagraph. The intensity scale follows the inverse hyperbolic sine.}
\end{figure*}

Following these steps the data were processed to remove the stellar halo (i.e. in order to achieve high-contrast) with the SpeCal pipeline (Galicher et al., in prep.) developed to reduce and analyse our GTO data from the SPHERE High-contrast Imaging Survey for Exoplanets (SHINE) survey. This pipeline implements a variety of ADI-based algorithms from which we used classical Angular Differential Imaging \citep[cADI,][]{marois2006}, Template Locally Optimized Combination of Images \citep[TLOCI,] []{marois2014}, and reference differential imaging (RDI) using a compatible reference star from our SPHERE GTO target library. 
In parallel, the IFS data were also processed using angular and spectral principal component analysis (PCA-ASDI) as described in \cite{mesa2015}. In the following we discuss the results based on the noADI (classical averaging with no subtraction), cADI (see Fig.~\ref{fig:CADI_IRD}), and RDI for the disk morphology. On the other hand, the photometric analysis of the point sources is based on TLOCI since it delivers more accurate photometry (Galicher et al., in prep.)

\subsection{IRDIS-DPI (BBH band)}  \label{sec:DPI obs}

The IRDIS-DPI observations of RY~Lup were carried out on 27 May 2016 with the BBH filter (1.625 $\muup$m, width= 290 nm) using the same apodized pupil Lyot coronagraph (N-ALC-YJH-S) as for the IRDIFS observations. Sixty-four polarimetric cycles were taken, each consisting of one data cube for each of the four half wave plate (HWP) positions (0, 45, 22.5, 67.5 degrees). Dedicated coronagraphic images were taken at the beginning and at the end of the science sequence to determine accurately the star centre behind the coronagraph using the four crosswise faint replicas created by the deformable mirror. 
The IRDIS data were first corrected for cosmetics using the DRH pipeline implemented at the SPHERE Data Centre and following the same steps as the DBI data. The outputs include cubes of left and right images (parallel and perpendicular polarized beams, respectively) recentred onto a common origin. 
The data were then reduced following the prescriptions of \cite{avenhaus2014}, using the azimuthal Stokes formalism (Q$_\varphi$, U$_\varphi$) and relative polarimetric measurements. The parallel and perpendicular polarized images are combined to produce Q and U Stokes images. To obtain clean Stokes Q and U images, that is to correct for instrumental polarization downstream of the HWPs position in the optical path, Q+ and Q- (0 and 45 degrees), and U+ and U- (22 and 67.5 degrees) are included in the image combination and used to compute the azimuthal Stokes components. The azimuth is defined with respect to the star position as shown by \cite{canovas2015} and the U$_\varphi$ signal is very small for centrally symmetric disks, but not for very inclined disks as shown by \cite{pohl2017}.

\begin{figure*}[hbtp]
  \centering
   \includegraphics[width=90mm]{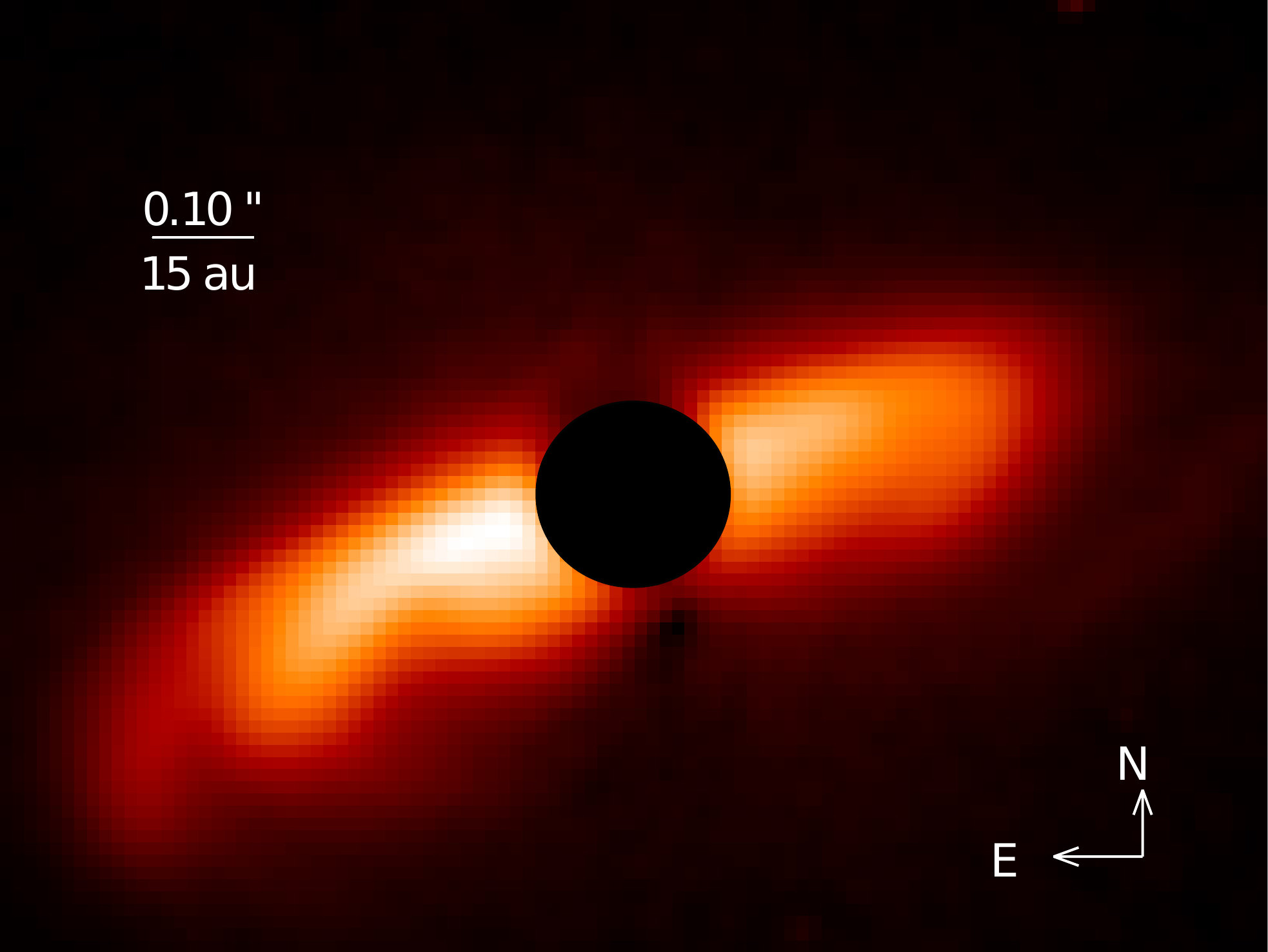}
  \includegraphics[width=90mm]{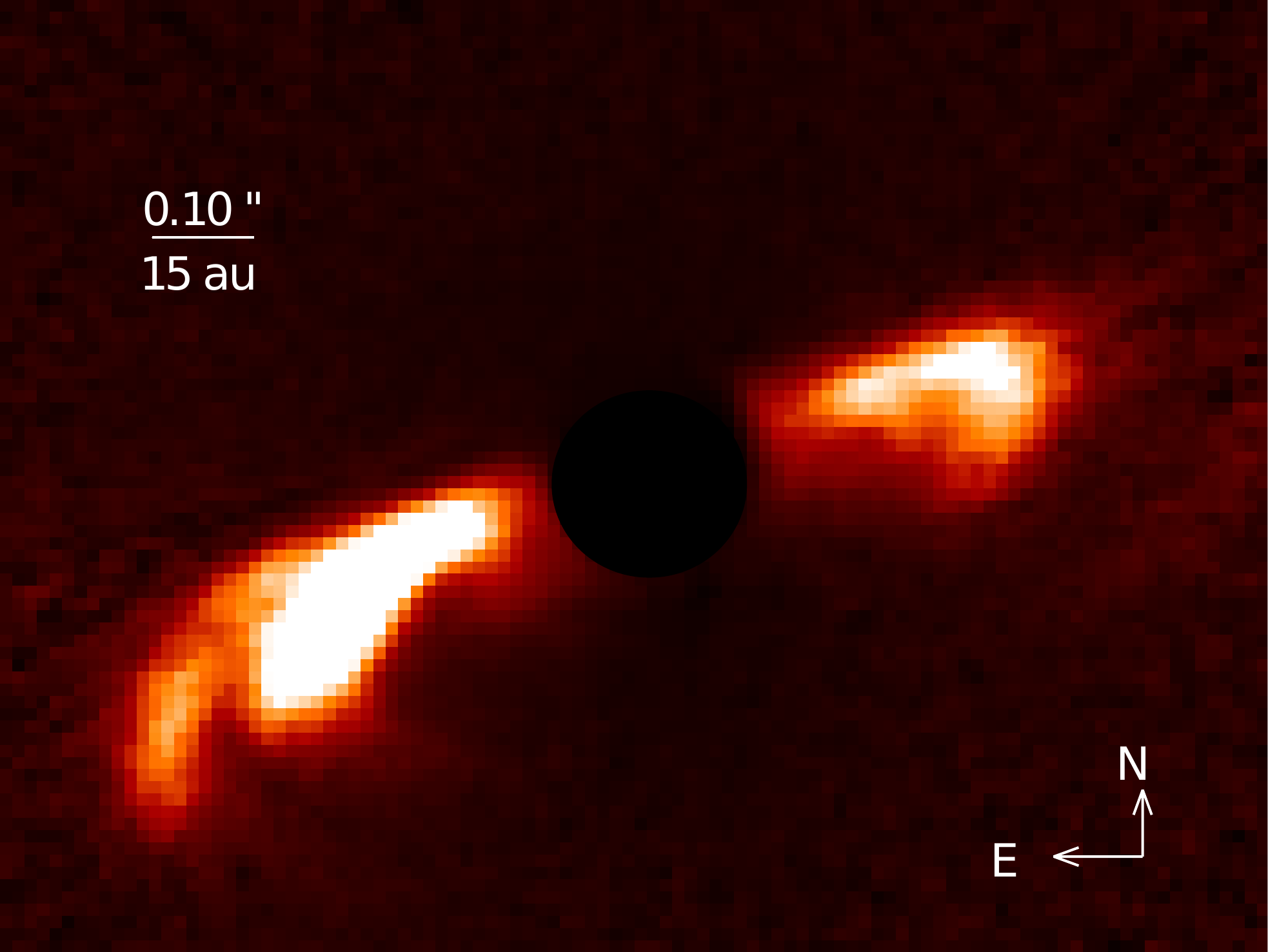}
  \caption[]{\label{fig:radial_DPI} (Top) Polarized intensity image showing the close-in environment of RY~Lup. (Bottom) Unsharp masking of the polarized intensity image showing the close-in environment of RY~Lup. The intensity scale follows the inverse hyperbolic sine. }
\end{figure*}

\begin{figure}
\centering
   \centering
  \includegraphics[width=100mm]{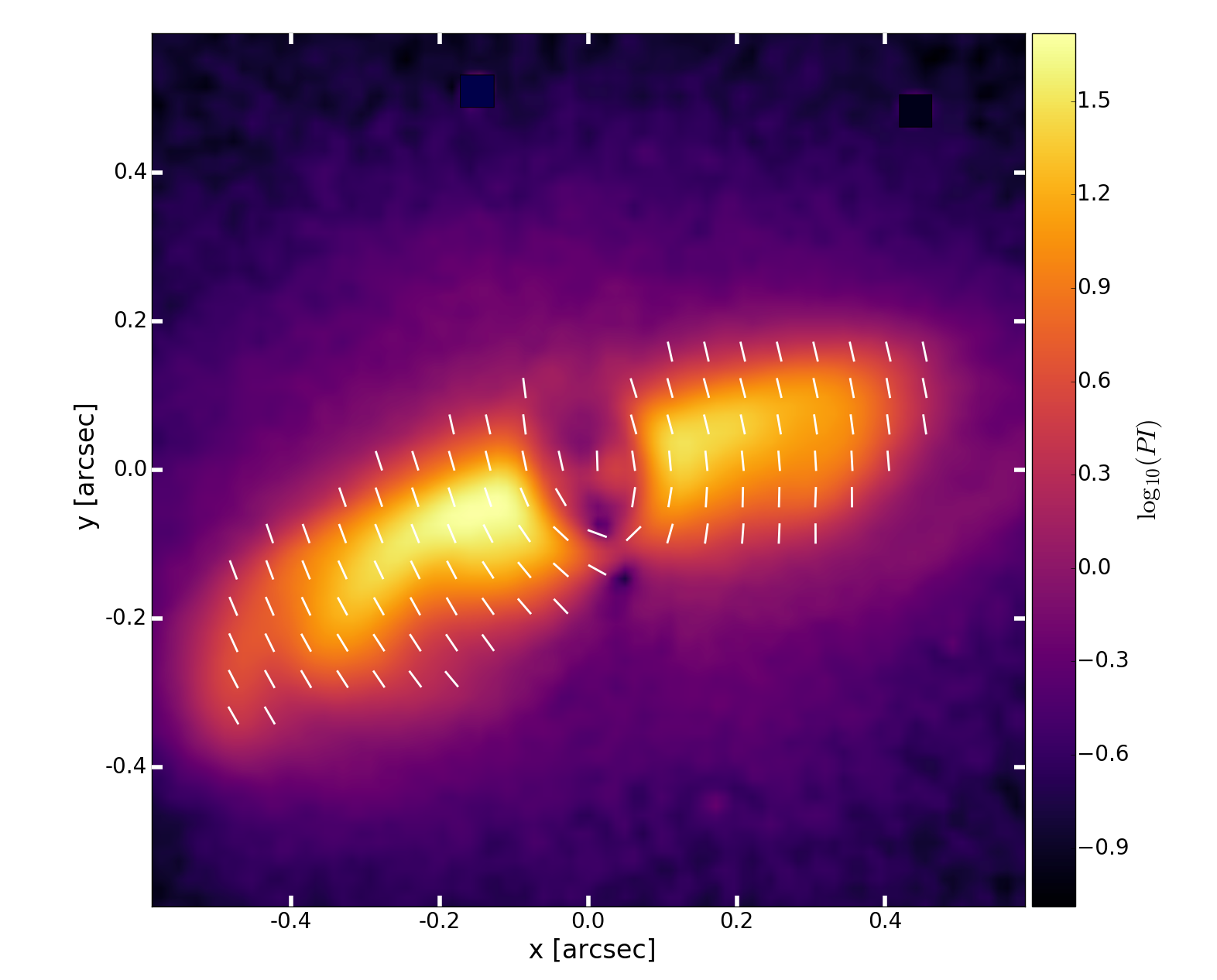}
  \caption[]{\label{fig:radial_DPI2} Polarized intensity image showing the close-in environment of RY~Lup. The white stripes represent the angle of linear polarization (fixed length, not scaled with the degree of polarization). The intensity scale follows the inverse hyperbolic sine. }
\end{figure}

However, there might still be an instrumental polarization left upstream of the HWP in the final Q and U images, which is assumed to be proportional to the total intensity image as shown in \cite{canovas2011}. Since the RY~Lup disk is inclined, the first standard reduction did not include U$_\varphi$ minimization of the signal in an annulus around the central star. We also reduced the IRDIS-DPI data with a new method based on an inverse approach method (Denneulin et al., (in prep.)). This method also enables the use of any set of frames with incomplete polarimetric cycles. The method is based on an optimization using the electric field (Jones Matrix) rather than the intensity (Mueller matrix) and relies on the inverse approach (i.e. fitting a model of the data to the observed dataset) which is significantly less biased and more efficient at minimizing instrumental artefacts. No assumptions about the angle of linear polarization of the source are made to correct for the instrumental polarization. The method is based on the use of an instrumental model describing the effect of the instrument on the incident electromagnetic field. The electric field is divided into non-polarized and polarized components, 
\[I_k = \Vert v_k \Vert^2  \rho + \vert \langle v_k, m \rangle \vert^2+ \text{noise},\] where $I_k$ is the flux measured on the detector, $\rho$ is the non-polarized intensity (unknown), $m$ is the electric field linearly polarized (unknown), and $v$ describes the effect of the instrument/telescope for a given half wave plate polarizer position. The model describes every polarimetric measurement and the polarized intensity is obtained by solving for both polarized and unpolarized components of the electrical field for every pixel individually using the  criteria
\[\min_{m, \: \rho} \quad \sum_{k} \frac{\omega_k}{2} \left\| \Vert v_k
\Vert^2  \rho + \vert \langle v_k, m \rangle \vert^2 - I_k \right\| ^2,\]
 where $\omega_k$ is a weighting factor calculated from the inverse of the noise variance.
The derived Q$_\varphi$ and U$_\varphi$ estimations using this latest method is similar to the standard reduction we have performed, but the estimation of the polarization angles is more accurate. In the RY~Lup case the disk is very bright which makes the level of instrumental polarization left in the Q$_\varphi$ and U$_\varphi$ images very small compared to the disk signal, and we conclude that the disk morphology is not affected by these polarimetric residuals. The Q$_\varphi$ and U$_\varphi$ images are displayed in Fig.~\ref{fig:pola_image} and Fig.~\ref{fig:radial_DPI}. Figure~\ref{fig:radial_DPI2} shows the polarized intensity overplotted with polarization vectors representing the angle of linear polarization. This strengthens the hypothesis that there is a clear departure from azimuthal polarization in inclined disks such as RY~Lup. 

 \section{Results}    \label{sec:Results}

We obtained a very clear detection of the RY~Lup disk in all datasets presented in this paper. The analysis of the morphology of the detected disk primarily focuses on the IRDIS and IFS intensity images and IRDIS-DPI (see Sects.~\ref{sec:IRDIFS obs} and \ref{sec:DPI obs}). Furthermore, the cADI IFS and IRDIS images are used to search for point-sources focusing on non-polarized companions because the intensity images reach higher contrast (Sect.~\ref{sec:planets}).  Our physical modelling of the polarimetric and total intensity images  allows a qualitative analysis of the disk, and is presented in Sect.~\ref{sec:diskmod}.

\subsection{Disk properties}

The IRDIS and IFS datasets reveal clearly the inclined disk around RY~Lup as seen in Fig.~\ref{fig:CADI_IRD}  in intensity and in Fig.~\ref{fig:pola_image} in polarimetry. Figure \ref{fig:CADI_IRD} for IFS is a median image over wavelength range from Y to J band. In our images, the disk appears as a dominating double-arch structure in the SE--NW direction extending to 0.53$\arcsec$  (80 \,au) in radius. Our observations support a high disk inclination with respect to the line of sight, with a position angle of 107 $\pm$1 degrees. The position angle was estimated by considering the brightest parts of the disk. In addition to these bright components the images reveal the presence of two fainter spiral-like features on the SE direction as seen on the display of the Q$_\varphi$ image (Fig.~\ref{fig:pola_image}). A spiral-like feature is also visible on the opposite side of the disk along its major axis. This either means that the $m=1$ and $m=2$ modes are basically seen on opposite sides of the disk or, together with the inner SE spiral arm, these two components could be approximately traced by an ellipse. In Fig.~\ref{fig:radial_DPI} we also applied an unsharp mask to the polarized intensity in order to enhance the disk structure. Other double-arch structures in TDs have been recently reported with high-contrast imaging instruments such as SPHERE \citep{janson2016, garufi2016,pohl2017}, but spiral-like features such as the one observed here have not yet been detected  in very inclined disks. The spiral-like features detected in the RY~Lup disk images could be created by an interaction of the disk with a planet located in its surrounding \citep{dong2016}. We  demonstrate in Sect.~\ref{sec:results} that it is a reliable hypothesis. 

We  note that the ADI processing of these images may have been biased and are not a faithful representation of the true intensity and geometry. In this particular case the disk is bright and these artefacts are negligible, as shown by the similarities between the cADI, noADI, and the DPI images. 

The reduced U$_\varphi$ image signal can be interpreted as radial polarization. The high peak-to-peak ratio between U$_\varphi$ and Q$_\varphi$ (30\%) can be attributed to the high inclination of the disk ($\sim$70 degrees as determined from the total intensity image).  In a highly inclined disk, multiple scattering can occur (i.e. where light that has already been polarized is scattered). In this case, this is the prime contributor to the U$_\varphi$ signal. This is consistent with a theoretical study by \cite{canovas2015} and \cite{pohl2017}, who found that multiple scattering can produce significant non-azimuthal polarization. They showed that the peak-to-peak ratio between U$_\varphi$ and Q$_\varphi$ can even be as high as 50 $\%$ for a disk inclination of 70 degrees depending on the mass and grain size distribution of the disk model. 

\begin{figure}[hbtp]
  \centering
   \includegraphics[width=93mm]{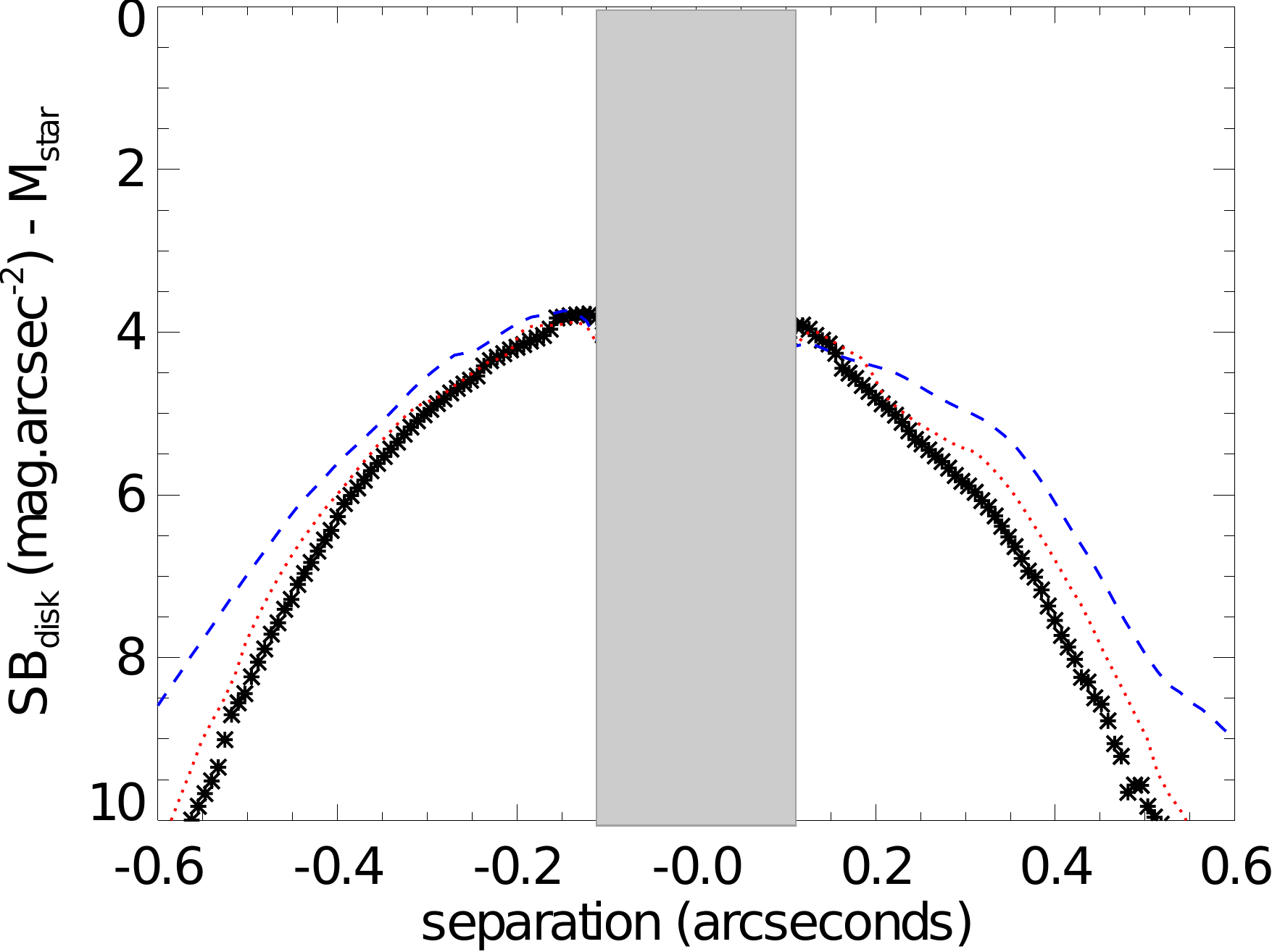}
  \caption[]{\label{fig:profile}
  RY~Lup disk surface brightness profile measured along the midplane on the IFS intensity image (asterisk line) expressed in $mag.{arcsec}^{-2} - M_{\star}$, IRDIS intensity image (red dotted line) expressed in $mag.{arcsec}^{-2} - M_{\star}$, and  IRDIS polarimetry image (blue dashed line) expressed in arbitrary units. The grey area represents the separation range masked by the coronagraph. The intensity is normalized to the unsaturated non-coronagraphic PSF intensity peak. The error bars are within 0.1 $mag.{arcsec}^{-2}$.}
\end{figure}

We used the RDI IFS, RDI IRDIS, and DPI IRDIS images to derive the relative surface brightness (SB-M$_{\star}$) profiles of the disk along its major axis (Fig.~\ref{fig:profile}), assuming that the major axis has a position angle of 107 degrees. These profiles are calculated using an average width of  5 pixels along the midplane with the following formula: SB-M$_{\star}$=$2.5*log$(normalized Intensity/${pixel-area}^2)- $M$_{\star}$. The conversion of the intensity into mag.${arcsec}^{-2}$ was performed using the 2MASS stellar magnitudes \citep[J = 8.54 mag, H = 7.69 mag,][]{cutri2003} and the ratio of the maximum to the total flux of the measured unsaturated non-coronagraphic PSF. With this normalization choice, the profiles provide information on the scattering efficiency of the dust grains. We note some brightness asymmetry between the west and east disk wings which is slightly more pronounced in the polarimetric intensity (represented with arbitrary surface brightness scale) than on the intensity cADI profiles (IFS and IRDIS). 
We also note that the radial profiles are very similar from the Y to H band, which suggests a shallow dependence efficiency of the scattering with wavelength. This is confirmed by the RY~Lup disk spectrum ranging from 5.05 to 5.25 $mag.{arcsec}^{-2} - M_{\star}$ across the 0.9 -- 2.3 $\muup$m wavelength range. This spectrum was estimated by computing the integrated flux from the IFS image at each wavelength in a constant area, defined by the area where the flux is above 1.5 $\sigma$ of the collapsed image over all wavelengths. 
The observed brightness and colour  of scattered light images put strong constraints on the scattering properties of protoplanetary dust as shown by \cite{mulders2012}. The very flat spectrum of the RY~Lup disk  suggests a minimum grain size of  $\sim$ 3 $\muup$m  assuming astrosilicates and a grain size distribution slope of  - 3.5. 

\subsection{Search for planets}
\label{sec:planets}

\begin{figure}[hbtp]
  \centering
  \includegraphics[width=90mm]{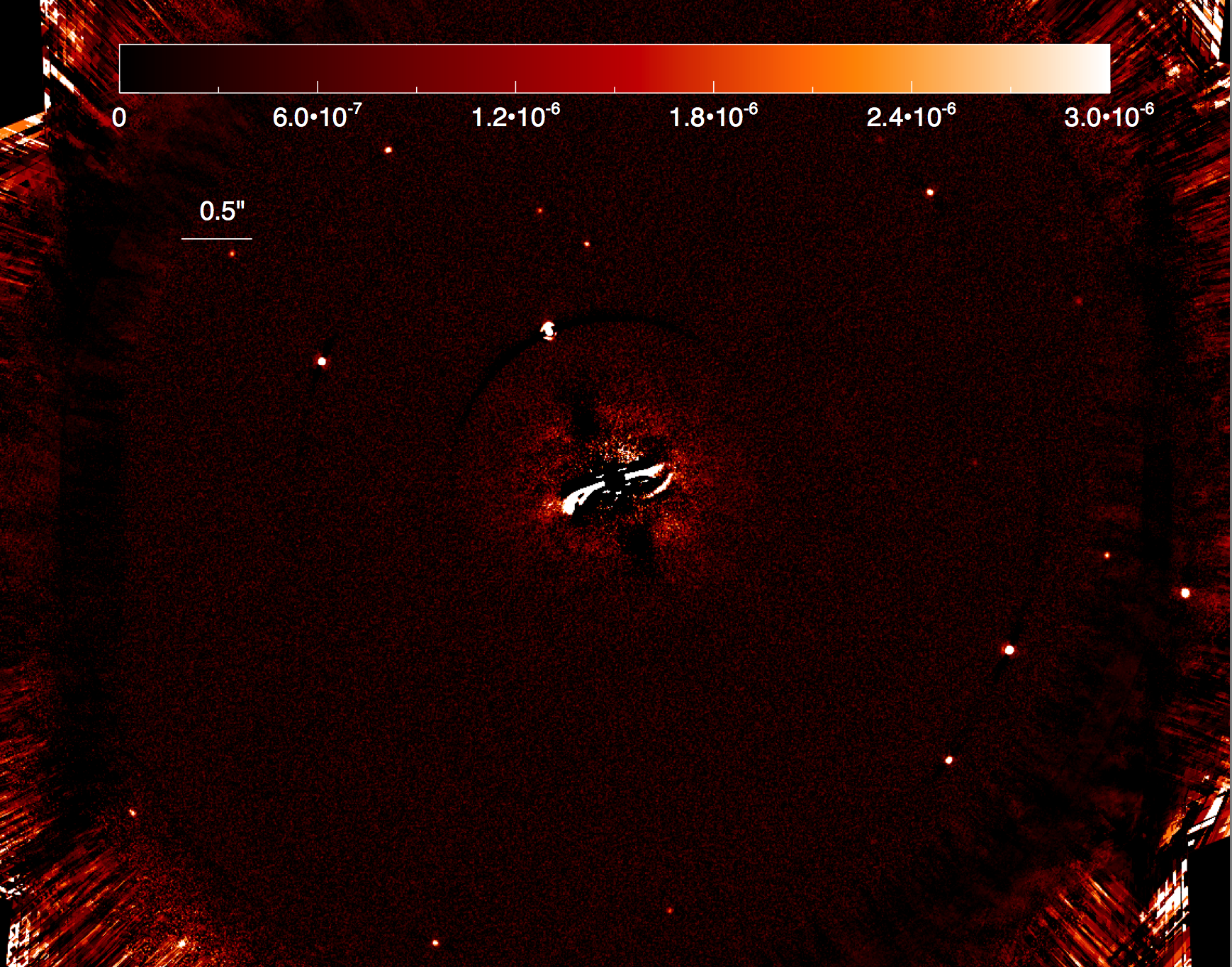}
  \caption[]{\label{fig:CC}
    IRDIS TLOCI H2+H3 image normalized to the maximum of the unsaturated non-coronagraphic PSF showing the detected point sources in the
environment of RY~Lup. The intensity scale is linear.}
\end{figure}

Twenty candidate companions (CC) were detected in the IRDIS field of view (Fig.~\ref{fig:CC}) using DBI with H2H3 filters, whereas no point-like source was found in the IFS image. The photometry and astrometry of the companions were measured using the TLOCI algorithm applied to each spectral band separately as presented in Table~\ref{table:CC}. We divided each science frame into annuli of 1.5 full width at half maximum  (FWHM, where FWHM = 41 mas). For each science frame and annulus, we computed the stellar residuals using the best linear combination of the most correlated frames for which the self-subtraction of mock point sources was at maximum 20\%. The astrometry and the photometry were estimated by a fit of the point spread function template onto the data. The photometric errors were estimated conservatively by including the variations in the stellar flux during the sequence (estimated from the fluctuations of the stellar residuals), the accuracy of the fitting procedure, and the PSF variability between the beginning and the end of the sequence. The astrometry of the detected companions was calibrated using pixel scales of 12.255 mas and a TN angle offset of -1.742 degrees \citep{maire2016}. The astrometric errors include the accuracy of the fitting procedure and the star centring error. 


\begin{figure*}
 \centering
    \includegraphics[width=80mm]{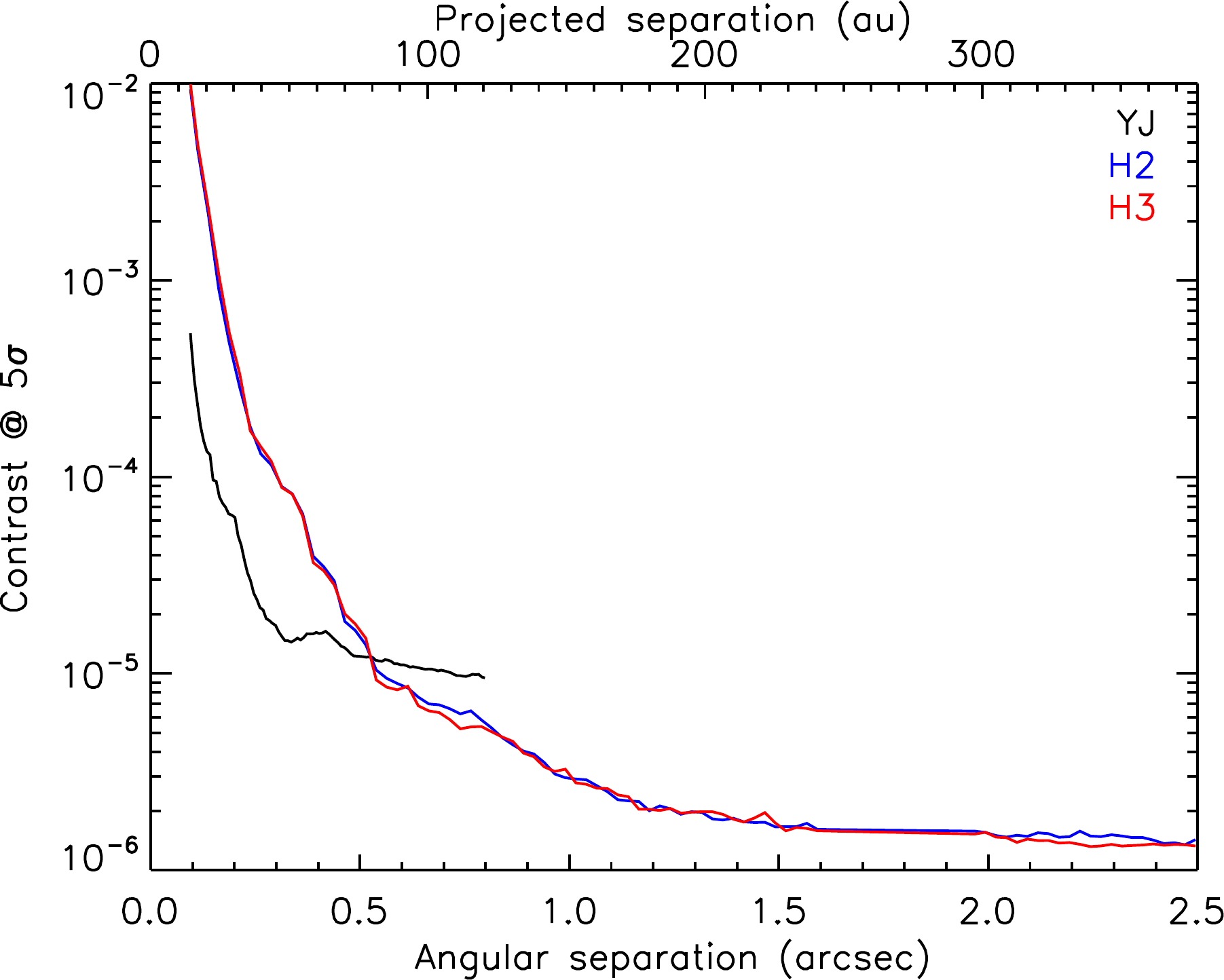}
    \includegraphics[width=80mm]{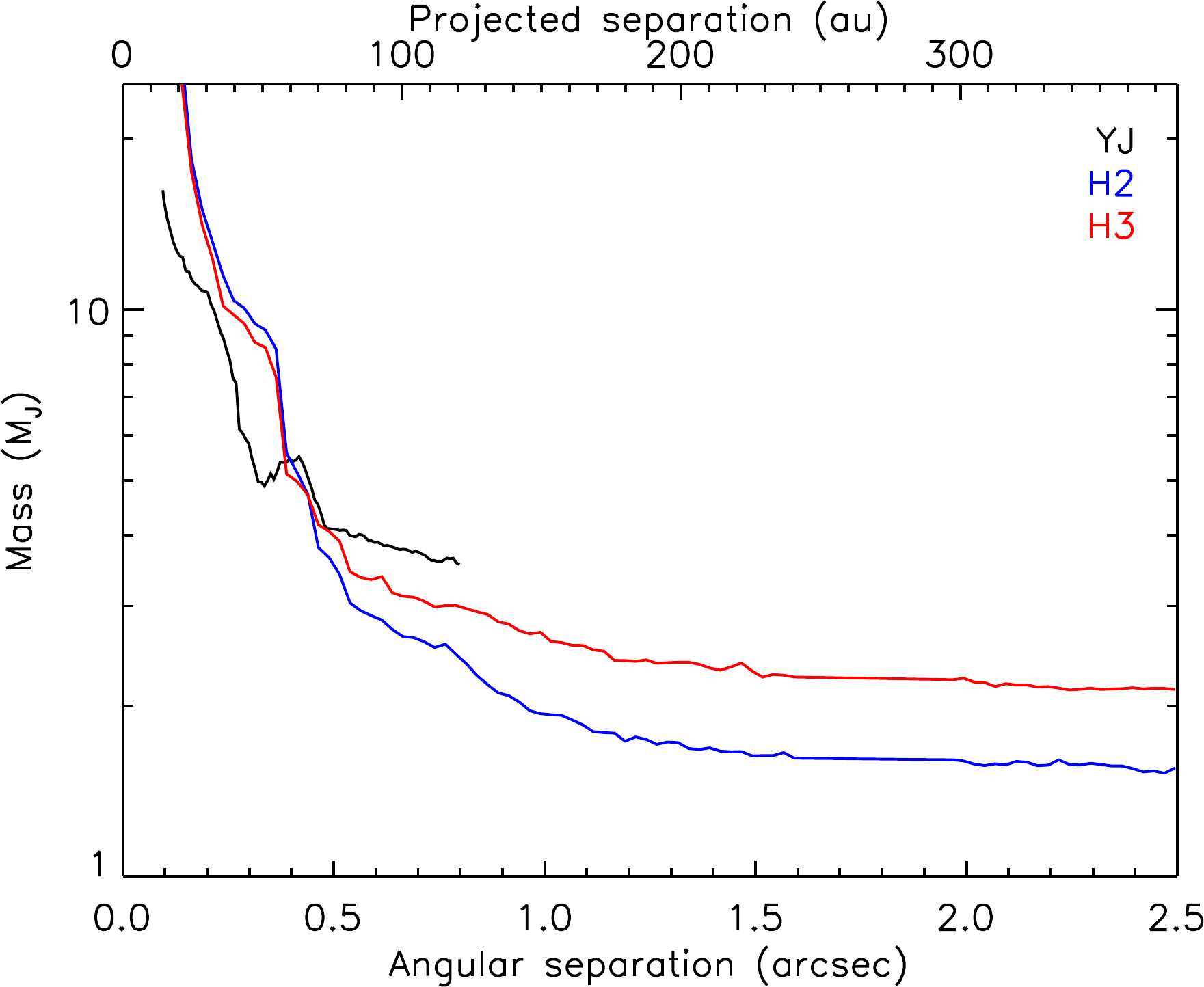}

  \caption[]{\label{fig:contrast}Contrast curves and companion mass limits derived for IFS (black curve) using the ASDI PCA reduction method, and for IRDIS H2 and H3 bands (blue and red curves, respectively) using the TLOCI reduction method. The coronagraph  inner working angle (radius where the coronagraphic transmission is 50\%)  is 0.095 arcsec in radius. }
\end{figure*}

\begin{figure}[hbtp]
  \centering
  \includegraphics[width=91mm]{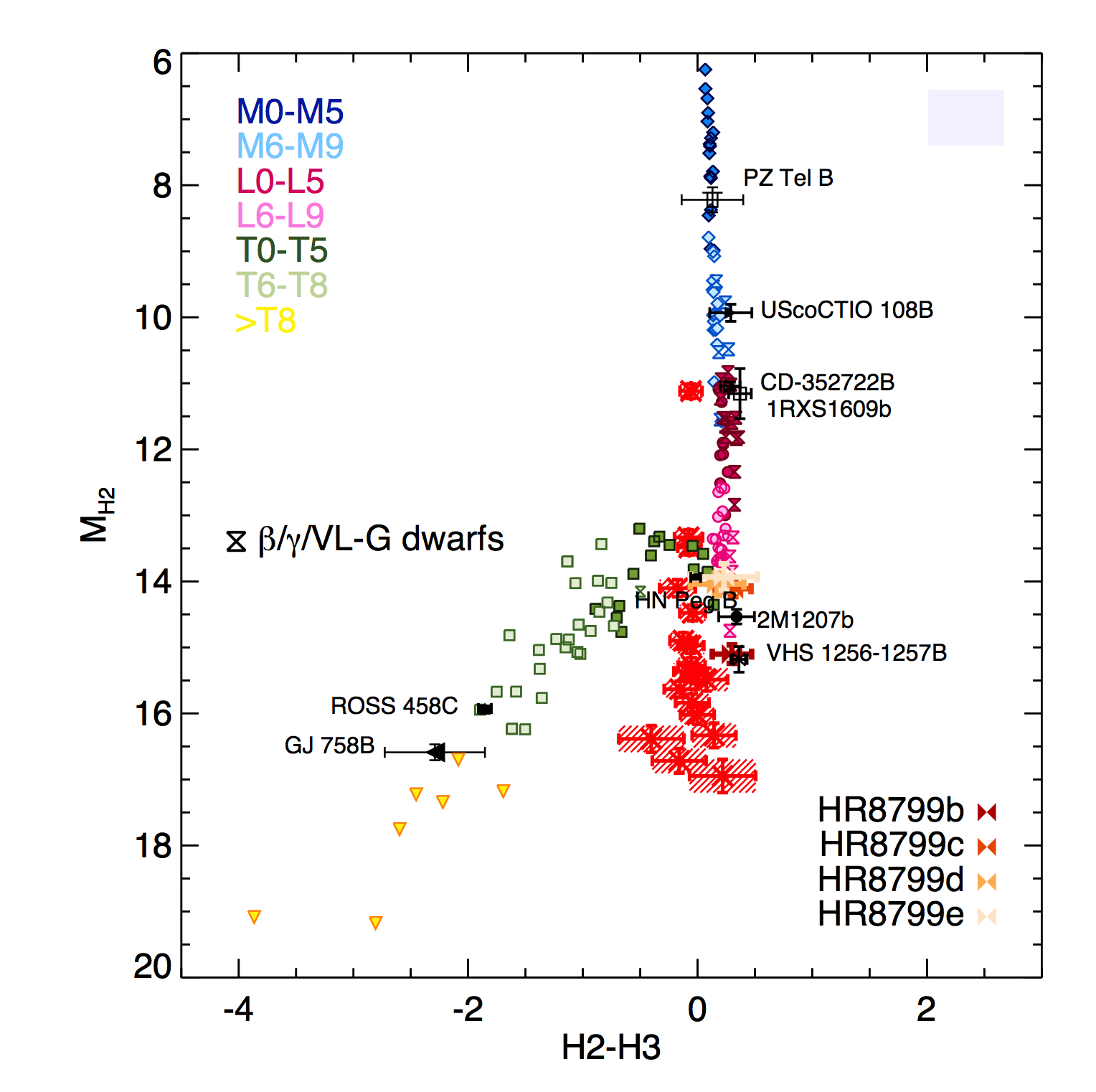}
  \caption[]{\label{fig:CMD}
    Colour magnitude diagram displaying the discovered candidate companions, which are marked in red including photometric error bars, compared to known substellar field (coloured symbols) and young objects. This plot assumes that the candidate companions are at the same distance as the star. The candidates were classified as probable background objects based on their position in the CMD. Several candidates lying near the L-T transition could still be considered as potential companions.
  }
\end{figure}

Because of the photometric bias from the TLOCI algorithm, we estimated the throughput of the technique at each position in the field  using fake planet injection and produced  throughput-corrected final images. The contrast curve for each spectral channel is estimated by the azimuthal standard deviation of the throughput corrected image in an annulus of 0.5 FWHM width. No small statistical correction at short separation was applied \citep{mawet2014}. The azimuthally averaged contrast curves shown in Fig.~\ref{fig:contrast} were estimated for the TLOCI reduction for IRDIS and PCA ASDI reduction for IFS because these reductions provide the best compromise for contrast, stellar rejection, and self-subtraction correction for the point source detection. The disk is bright, but strongly attenuated by the TLOCI algorithm (which is optimized for point-source detection). As a consequence, the disk residuals increase the measured noise level by a small amount leading to slightly pessimistic contrast curves at very short distances. For the conversion of the contrast limits to mass limits in Fig.~\ref{fig:contrast}, we used the COND atmospheric and evolutionary models of \cite{baraffe2015}  and \cite{baraffe2003}. All the point sources were classified as probable background stellar objects using the SPHERE consortium tools developed for the classification and the ranking of the companion candidates discovered in SPHERE/SHINE. With the same suite of tools, we derived the colour magnitude diagram (CMD) of all point sources, as shown in Fig.~\ref{fig:CMD}, to compare their H2-H3 colour to those  of field and young dwarfs covering the M, L, T, and Y types. Further details about the derivation of this colour magnitude diagram are provided in \cite{zurlo2016}. We also computed a background probability for the point sources by assuming the Besancon model predictions \citep{robin2003}. As seen in Table~\ref{table:CC}, there is only one candidate with low probability of contamination, but its position on the CMD diagram corresponds to a background star. All the detected point sources also have   very similar offsets from the M-L sequence and their separations are also greater than a hundred \,au, so we can conclude that they have a very low likelihood of  being bound companions.

 \begin{table*}[t]
\caption{Astrophotometric parameters for the CCs. For Candidates 14 and 19 proper astrophotometric measurements could not be extracted at one of the two wavelengths.The last column gives the contamination probability}
\label{table:CC}      
\centering                                      
\begin{tabular}{c c c c  c c c c c c}          
\hline\hline                        
 CC \# & Sep (mas) &Err Sep (mas) & PA (deg)  &Err PA (deg) & Mag H2 & Err Mag H2 & Mag H3& Err Mag H3 & prob (\%) \\    
\hline\hline                               
  
  \textbf{1}         &   5047.88 &     6.51 &     59.28    & 0.14&13.84&    0.09&  13.99&     0.11 &98.5 \\ 
     \hline      
       \textbf{2}        &    3587.47    &  3.89    &  67.75  &   0.13&   11.69 &   0.07 &11.76 &   0.071&60.7\\ 
     \hline      
      \textbf{3}       &     1855.11  &    2.60   &   23.80&     0.13& 9.32 &   0.07 &9.38 &   0.067 &5.8\\ 
     \hline      
     \textbf{4}        &  2704.75     & 4.29    &  6.80    & 0.14 &13.63&    0.08&13.66&    0.08 &68.9 \\
     \hline      
      \textbf{5}       &  4078.96    &  10.53   &   273.02 &    0.19 &14.92&     0.16& 15.08    & 0.17 &97.4\\ 
     \hline      
     \textbf{6}       &  5633.39    &  9.07  &    291.21 &   0.15 & 14.54&     0.16& 14.40&   0.10&99.8\\ 
     \hline      
       \textbf{7}      &    5632.97  &    6.45   &   261.47  &   0.13 &  13.57    &0.09 &13.65  &  0.09&99.2 \\ 
     \hline      
       \textbf{8}     &     6579.01   &   9.63   &   258.92  &   0.14& 12.31&    0.10& 12.49&     0.12 &98.5\\ 
     \hline      
      \textbf{9}     &    4855.73   &   5.59  &    246.86  &   0.13 & 11.54&    0.08& 11.62  &  0.10&79.1 \\ 
     \hline      
      \textbf{10}      &    4545.61   &   5.30  &    34.42   &  0.13& 13.18&   0.08&  13.24&    0.084 &93.9\\ 
     \hline      
      \textbf{11}      &     3182.06  &    6.02  &    15.52   &  0.16& 14.05    & 0.10 & 14.09  &   0.10 &83.9 \\ 
     \hline      
      \textbf{12}      &    4896.42   &   12.88   &   322.20  &   0.19&15.15&     0.24& 14.93&     0.16 &99.6 \\ 
     \hline      
       \textbf{13}   &    4840.22    &  5.53    &  312.53  &   0.13& 13.10& 0.08& 13.23 &   0.08&95.2\\ 
     \hline      
      \textbf{14}   &    5782.48    &  23.59     & 298.59    &     0.26&15.59 &   0.61&   &&99.9 \\ 
     \hline      
      \textbf{15}   &     4922.87    &  5.096   &   230.19  &  0.13&12.68  &  0.08& 12.73&    0.08 &93.0 \\ 
     \hline      
      \textbf{16}   &        6626.06  &    8.87   &   124.44   &  0.14 & 13.69 &  0.15 & 13.62 &0.11&99.9\\ 
     \hline      
      \textbf{17}     &      5603.79    &  6.02   &   158.70  &   0.13 &13.48  &  0.08& 13.54&    0.081&99.1 \\ 
     \hline      
      \textbf{18}    &       4892.20    &  6.83   &   187.26  &   0.14 & 14.23&     0.113 & 14.23&    0.093&98.9\\ 
     \hline      
       \textbf{19}    &        7173.35   &   9.50    & 136.834     &0.14& &&13.01&     0.12  &99.8\\ 
     \hline      
       \textbf{20}      &     6639.75    &  12.95   &   79.77   &  0.16& 14.59    & 0.18 &14.99&     0.22 &99.9\\ 
   \hline \hline
\end{tabular}
\end{table*}


\section{Physical disk modelling}
\label{sec:diskmod}

Our physical disk modelling aims to reproduce the polarized intensity, the orientation of the polarization vectors, and the two faint spiral arms observed on the SE and SW sides in the SPHERE intensity and polarimetric images. This requires the loss of disk symmetry, which could be explained by planet--disk perturbation dynamics. Thus, we want to test whether the spiral pattern triggered by a Jupiter-like planet could eventually result in the scattered light features that we detect around RY~Lup, assuming that its disk is seen close to edge-on. For this purpose, we carry out 3D global numerical hydrodynamical simulations to study the morphology of spiral arms excited by an outer planetary perturber. Using detailed 3D follow-up radiative transfer models, we produce synthetic scattered light images for a highly inclined disk configuration at NIR wavelengths. This modelling approach is motivated by recent work on spiral arms in the context of scattered light by \citet{dong2015,dong2016,juhasz2015,pohl2015}.


\subsection{Hydrodynamical planet--disk simulations}
\label{subsubsec:hydro}

We use the  Three-dimensional RAdiation-hydrodynamical Modelling Project (TRAMP, \citealt{klahr1999}) code in its extended version (\citealt{klahr2006}), which applies a flux-limited diffusion approximation for the radiation. We consider a giant planet embedded in a fully viscous disk, where the consistent release of accretion energy from the planet via radiation is included. The disk model considers a spherical polar coordinate system ($r,\theta,\varphi$), where the disk midplane coincides with the $\theta=0$ plane. The number of grid cells in the radial ($r$), polar ($\theta$), and azimuthal ($\varphi$) directions are 64, 31, and 89, respectively. The computational disk grid has a radial extent from $\sim$18\,au to 235\,au. The complete set of vertical disk structure equations is solved self-consistently, including both viscous dissipation and heating by irradiation from the central star (cf. \citealt{dalessio1998}). This results in detailed profiles of the density and temperature structure with vertical height and disk radius. Thus, it allows us to determine the 3D temperature structure of the disk at each step of the planet--disk  interaction processes. More precisely, the initial 3D dataset, before adding the planet, is determined from a set of 1D vertical structure models for a given accretion rate (which will result in the desired total disk mass). Thus, the dynamical viscous evolution in the 3D radiation hydro simulation of the disk will not  significantly alter the initial global gas distribution, but any development of structure in the 3D simulations is the result of the torques exerted by the planet. We make sure that the 1D and 3D models use the same parameters including the dust opacities and alpha viscosity. We assume a central stellar mass of $M_{\star}=1.4\,M_{\odot}$, a radius of $R_{\star}=1.72\,R_{\odot}$, and an effective temperature of $T_{\mathrm{eff}}=5669$\,K. These parameters are consistent with the properties of RY~Lup derived by a piecewise linear interpolation of the values from \citet{manset2009} assuming an object's distance of 151\,pc \citep{gaia}. Furthermore, the input parameters for the viscosity parameter and the mass accretion rate are chosen to be $\alpha=10^{-3}$ and $M_{\mathrm{accr,0}}=3.3 \cdot 10^{-10}\,M_{\odot}\,yr^{-1}$, respectively. The latter is rather low compared to typical values for T Tauri disks, but a natural outcome of the disk mass considered. The accretion rate is calculated such that the disk gas mass corresponds to $M_{\mathrm{disk}}=2.5 \cdot 10^{-3}\,M_{\odot}$, distributed between an inner disk radius of $r_{\mathrm{in}}=18\,$au and an outer disk radius of $r_{\mathrm{out}}=123\,$au, as measured by \citet{ansdell2016}.

The position of a possible planetary perturber is uncertain, but spiral arms exterior to a planet's orbit are unlikely to explain the observations as they are too tightly wound given typical disk scale height values \citep[][]{juhasz2015}. However, a location further out in the disk is compatible with the outer disk radius in the $^{13}$CO ALMA emission map. Furthermore, the mm dust continuum suggests a truncation at $\sim$120\,au and a depletion of large dust beyond \citep{ansdell2016}. Following the approach by recent spiral arm models \citep[][]{dong2015}, the planet can be roughly located at a distance three times the location of the inner arms. Thus, we consider a planet located at 190\,au, that is outside of the disk radius in both scattered light and in thermal emission. Given that the planet carves out a wide gap in the disk, this location is a justified assumption. We assume a fixed circular orbit, although spiral density waves may be also excited by a companion on an eccentric orbit. The mass of the embedded giant planet is considered to be M$_{\mathrm{pl}}$/M$_{\star}=1.4 \cdot 10^{-3}$ ($\sim$2\,M$_{\mathrm{jup}}$). This mass is restricted by the mass detection limits for planetary companions obtained from our SPHERE contrast curves. Figure \ref{fig:contrast} shows that the upper mass limit at an angular distance of $\sim$1.1\arcsec is $\sim$2\,M$_{\mathrm{jup}}$. It is worth noting that the mass limit is model dependent and slightly higher if a warm start model is assumed ($\sim$5\,M$_{\mathrm{jup}}$) \citep[]{marley2007, Spiegel2012}. It is also worth mentioning that a planet might be at a location behind the disk or might be geometrically close to the central star, where it could remain undetected. Hence, it is geometrically likely to have a projected separation less than $\sim$1.1\arcsec, at which a higher planet mass would be possible according to Fig.~\ref{fig:contrast}. However, our choice of a lower planet mass is thus conservative, which is reasonable given the uncertainties on the age of the system and on the atmospheric/evolutionary models used to estimate the mass detection limits. The planet--disk features, especially the contrast between spirals and the background disk, would be even stronger for higher planet masses \citep[cf.][]{dong2015,pohl2015}. The simulation is run for 90 orbits at 190\,au ($\sim 1.8 \cdot 10^5$\,yrs), which is sufficiently long for the inner spiral arms to reach a quasi steady state, although this evolutionary timescale would be not enough to carve out a full gap around its orbit. This is, however, reasonable for our modelling purposes since the planet is located outside the disk and we are mainly interested in the observational appearance of the spiral arms.

 \begin{figure}
        \centering
        \centerline{
                \includegraphics[width=\columnwidth]{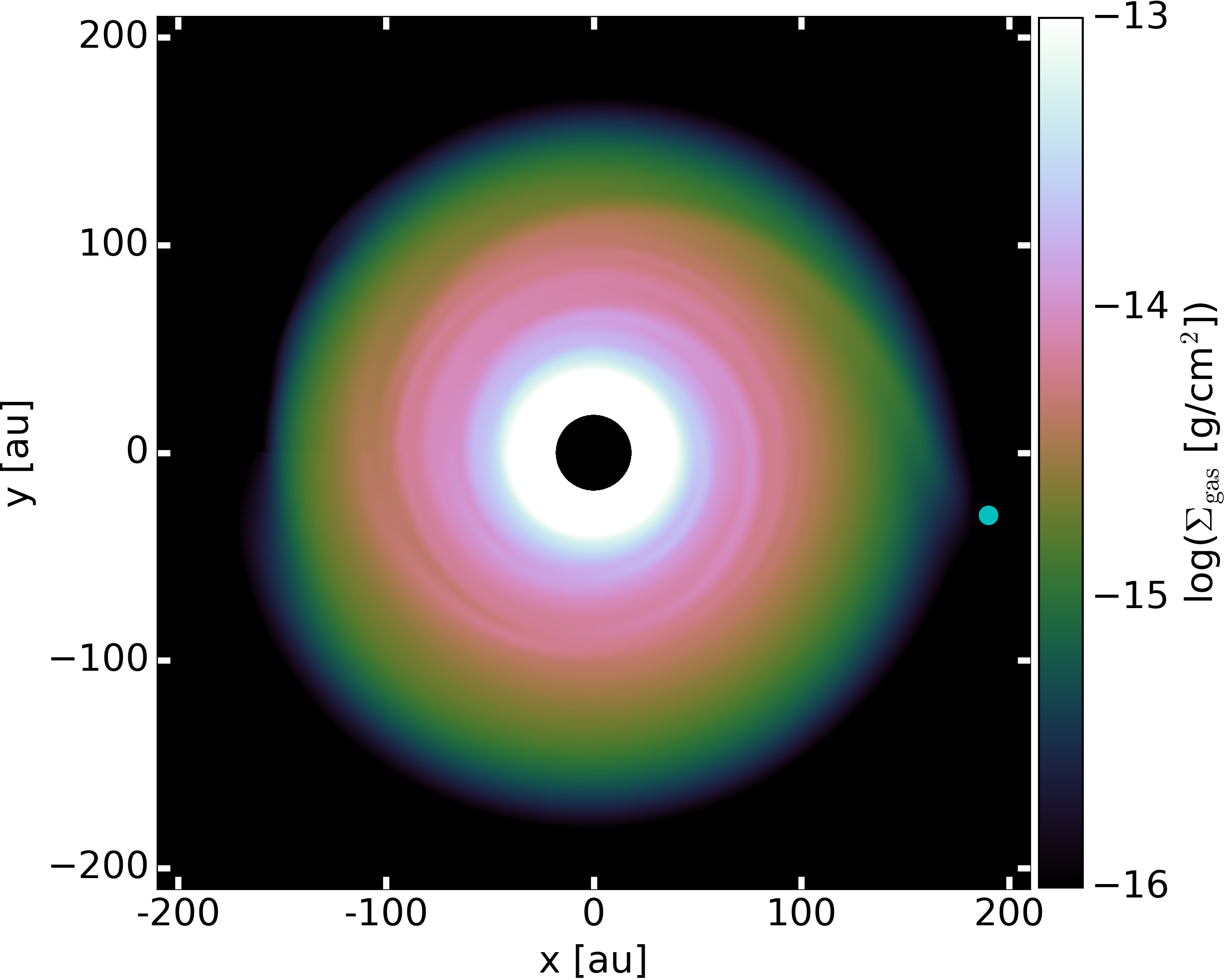}
        }
        \caption{Simulated 2D vertically integrated density map for a planet-induced spiral wave considering a planet-to-star mass ratio of $1.4 \cdot 10^{-3}$. The green dot highlights the position of the planet.}
        \label{fig:2mjup_density_tramp}
\end{figure}

A map of the surface density $\Sigma_{\mathrm{gas}}$ after an evolutionary time of 90 orbits at 190\,au is shown in Fig.~\ref{fig:2mjup_density_tramp}. While the planet always excites inner and outer waves, we concentrate here on the $m=2$ spiral arm configuration inward of the planet's position at 190\,au. The primary arm originates from the planet location, the secondary is shifted by $\sim$180$^{\circ}$ in the azimuthal direction. The density contrast of the secondary spiral wave with respect to the background disk is weaker than the primary one.

\subsection{Radiative transfer models}
\label{subsubsec:rtmod}

The resulting 3D disk density and temperature structure presented in Sect.~\ref{subsubsec:hydro} is subsequently fed into the Monte Carlo (MC) radiative transfer code \textsc{RADMC-3D} developed by \citet{dullemond2012}\footnote{The \textsc{RADMC-3D} source code is available online at http://www.ita.uni-heidelberg.de/$\sim$dullemond/software/radmc-3d/.}. To convert the gas density to the dust density used in the radiative transfer calculations, we adopt a gas-to-dust ratio of 50, which is lower than the canonical ISM value as suggested by the ALMA Lupus survey of \citet{ansdell2016}. For the dust opacity calculation we consider a species mixture of silicate (60\%), carbonaceous (10\%), and icy (30\%) material, where the optical constants are taken from \citet{draine2003,zubko1996,warren2008}. We assume two different grain size populations, small (0.01--1\,$\mu$m) and large (1\,$\mu$m--1\,mm) grains, where each size distribution is a smooth power law with exponent $-3.5$. The mass ratio between the two populations is determined such that the number density follows $n(a)\,\propto\,a^{-3.5}$. Thus, the large fraction of mass is in the large grains ($\sim$95\% of total dust mass), but the small grains dominate the opacity at NIR wavelengths. We assume that the gas and the small dust is well mixed, thus their dust scale height is equal to the pressure scale height. We are aware that this choice does not reproduce the scattering colours of the disk, but we only focus on the morphology of the image in our model approach.

The stellar parameters for the radiation source as well as the radial disk extension are taken to be the same as in the hydrodynamical calculations. For the polar and azimuth we use a finer grid sampling with N$_{\varphi}=192$ and N$_{\theta}=64$, and interpolate the density and temperature array values accordingly. All scattering MC simulations are run with $5 \cdot 10^{8}$ photon packages. We consider multiple scattering effects and produce images for all four Stokes components, from which the total intensity and polarized intensity images are calculated at \textit{H} band. The disk orientation is set by the inclination angle ($0^{\circ}$ is face-on) and the position angle (PA, from north to east). The theoretical radiative transfer images are convolved by the instrument PSF with a FWHM of 0\farcs04 so that it mimicks the angular resolution of the data.

\section{RT Modelling results}
\label{sec:results}
\subsection{Synthetic scattered light images}
\label{subsec:results_scattered}

\begin{figure}
        \centering
        \centerline{
                \includegraphics[width=\columnwidth]{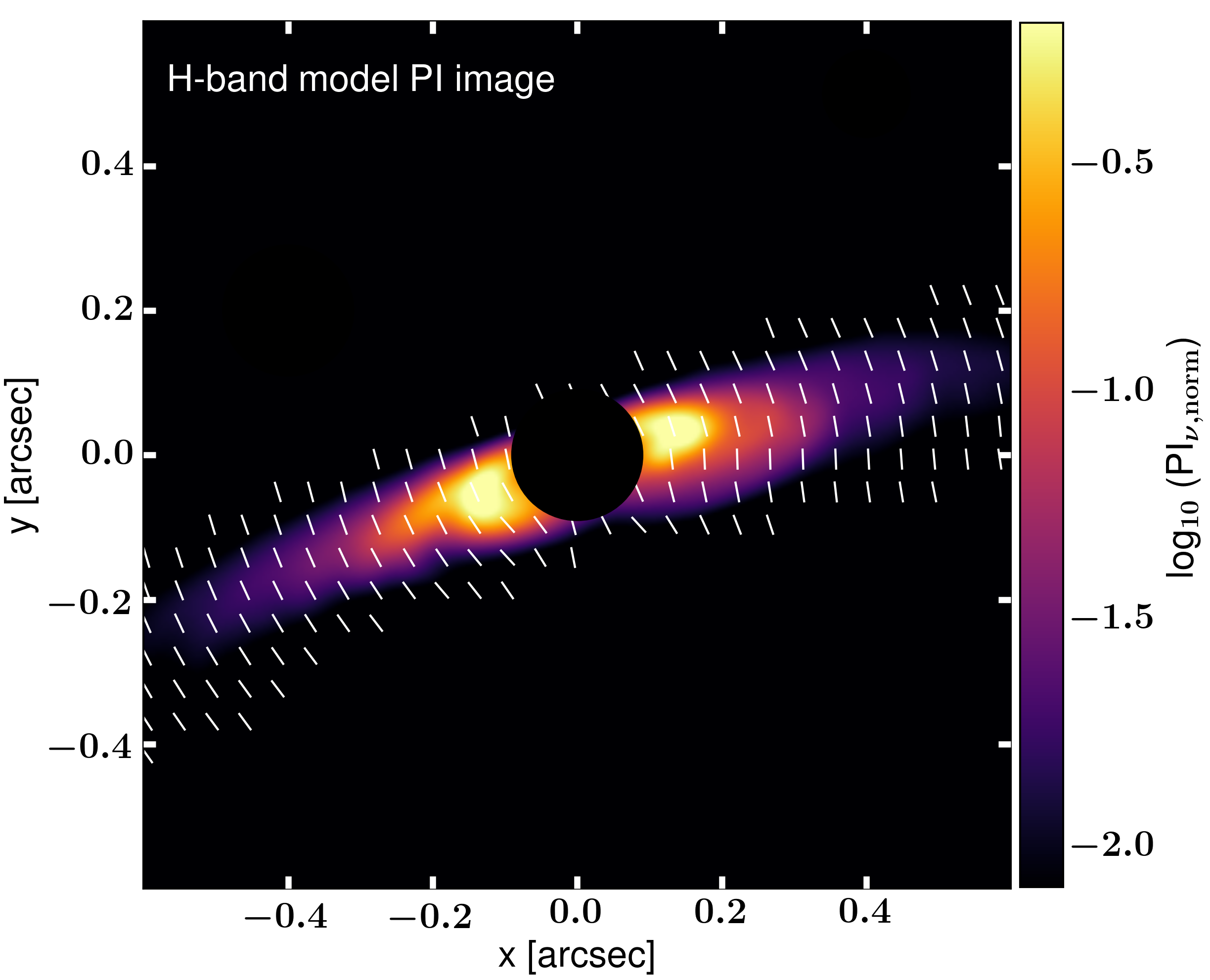}
        }
        \caption{Convolved polarized intensity \textit{H}-band image considering a planet-induced spiral model with M$_{\mathrm{pl}}$/M$_{\star}=1.4 \cdot 10^{-3}$. The image is produced at an inclination of 75\,deg, and a PAs of 107\,deg. The inner $0.18''$ are masked (black circular area) to mimic the coronagraph region. The white stripes represent the angle of linear polarization (fixed length, not scaled with the degree of polarization).}
        \label{fig:2mjup_pi_conv}
\end{figure}

\begin{figure}[hbtp]
  \centering
  \includegraphics[width=91mm]{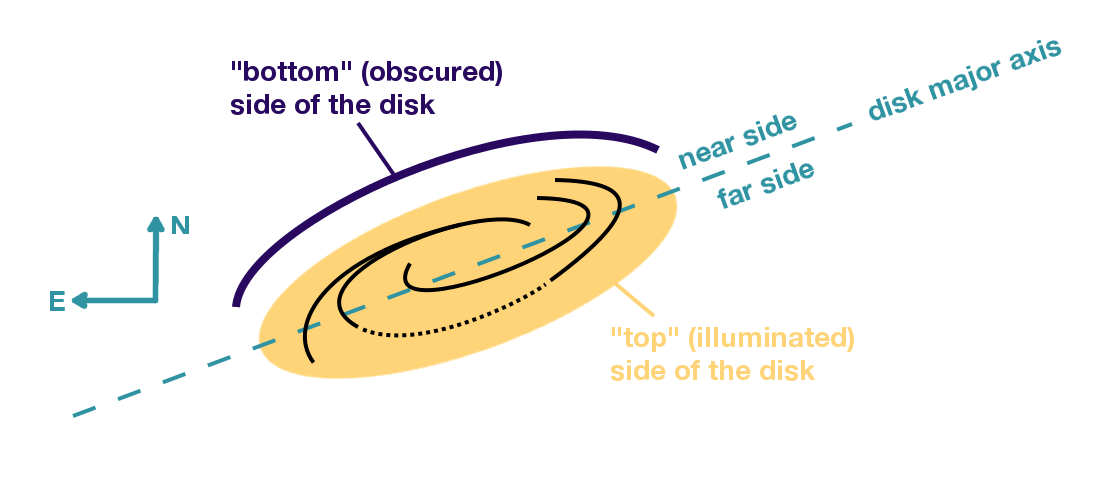}
 \caption{Sketch of the disk model suggested by our observations and modelling. North is up and east is towards the left. The major axis of the disk is indicated by the dashed line and divides the disk into a near and far side. The top (illuminated) and bottom (obscured) halves are separated by the disk midplane.}
  \label{fig:sketch}
\end{figure}

The convolved H-band polarized intensity image for our planet--disk interaction model in the RY~Lup system is shown in Fig.~\ref{fig:2mjup_pi_conv}. The disk is inclined by 75\,deg with respect to the line of sight and rotated by a PA of 107\,deg relative to the disk major axis. We note that we ran a grid of models for inclinations between 70 and 80\,deg in steps of 1\,deg, where the data is best represented by an inclination of 75\,deg. This is consistent with the measurement from total intensity given the uncertainties. For such a highly inclined system, the symmetric two-spiral arms produced by the external planetary perturber and known from face-on disk configurations can have a completely different morphology in scattered light \citep[cf.][]{dong2016}. The spiral arms ($m=1$ and $m=2$ modes) instead appear as two arms located along the major axis at the top (illuminated) side of the disk (see sketch in Fig.~\ref{fig:sketch}). There is also a substructure branching off the SE spiral, possibly a second winding of the spiral density wave, but in the convolved synthetic image it is not as pronounced as in the observed PDI image. Both spiral arms have an almost equal brightness contrast in the scattered light image. This result is consistent with a parametric study by \citet{dong2016} on how spirals driven by companions can appear in scattered light at arbitrary viewing angles.

\subsection{Polarimetric comparison}
Our disk model leads to a qualitatively good match with the IRDIS polarimetric image from Fig. \ref{fig:pola_image}. The bright double-arch structure in the SE--NW direction extending to 0.53$\arcsec$ in radius to the NW is well reproduced and corresponds to scattered light from the bottom side of the disk. The model also confirms that the disk must have a high inclination with respect to the line of sight. In addition to these bright components the two fainter spiral-like features in the SE direction, which break off, are also reproduced. Due to the multiple disk and planet parameters to adjust in these models and due to the high inclination of the disk, there is no perfect match of the brightness ratios between the east and west sides, and the bright and fainter features of the disk. For such a highly inclined system, the symmetric two spiral arms generated by the external planetary perturber in fact have very different morphology in scattered light with the viewing angle.

Figure~\ref{fig:2mjup_pi_conv} shows our polarized intensity model for RY~Lup overlaid with linear polarization vectors. We note that all vectors have the same length as we are especially interested in the polarization orientation. As also detected in the IRDIS polarimetric image (Fig.~\ref{fig:radial_DPI}), it is noticeable that the polarization direction along the minor axis is radial, which  means that  there is a significant non-azimuthal polarization. Since the polarization angle is indeed more complicated than expected for single scattering, the $U_{\varphi}$ signal must be physical and connected to multiple scattering, which is included in our radiative transfer calculations as well.



\subsection{Spectral energy distribution}
\begin{figure}[hbtp]
  \centering
  \includegraphics[width=91mm]{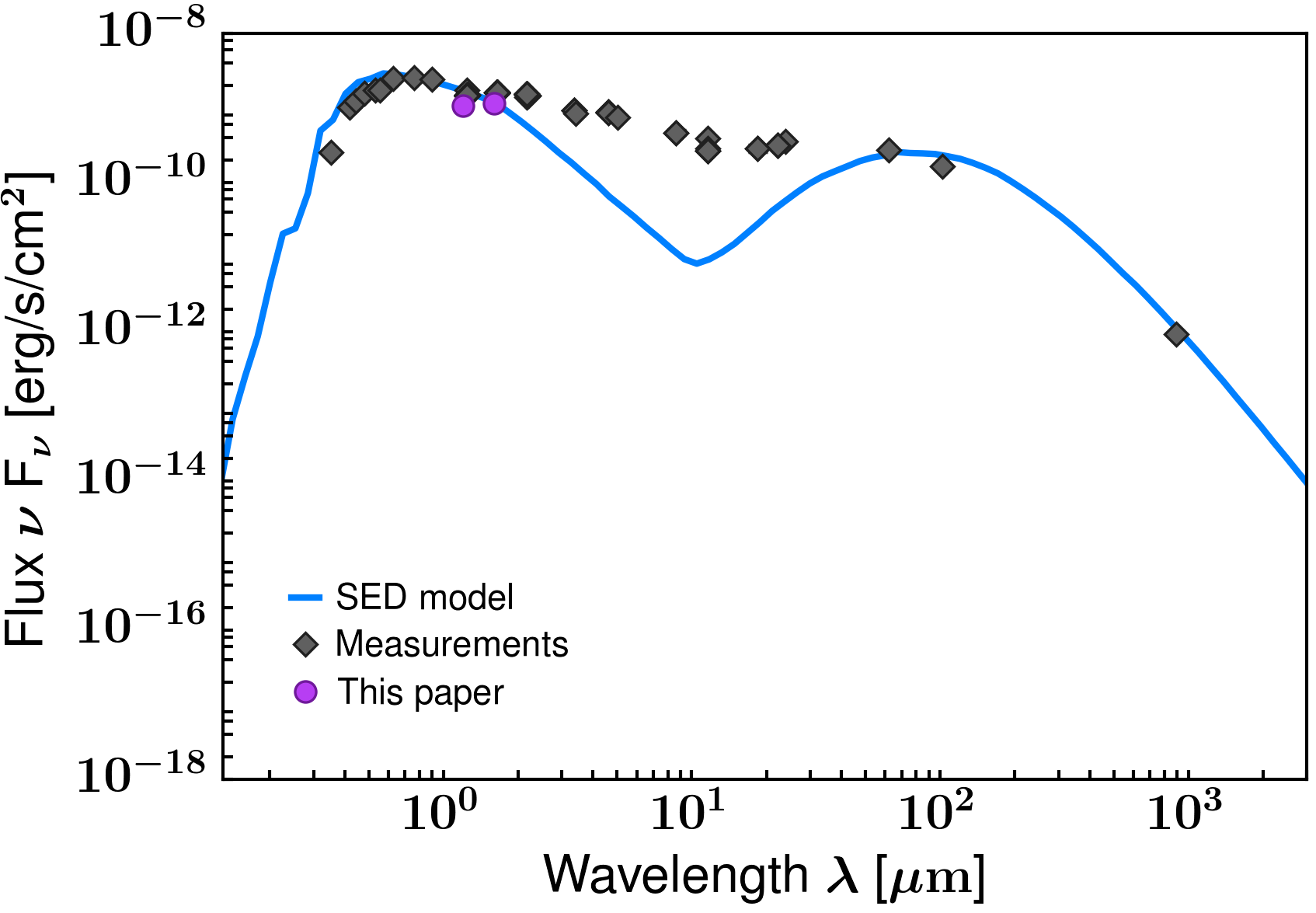}
  \caption[]{ Synthetic SED from our radiative transfer model compared with direct measurements obtained with SPHERE IRDIS and IFS in violet, VLTI/PIONIER (Anthonioz 2015), 2MASS, IRAS, WISE, and AKARI (VizieR online catalogue collection\footnote{http://vizier.u-strasbg.fr/vizier/sed/}) and ALMA \citep{ansdell2016} in grey.\label{fig:SED}
  }
\end{figure}

The SED model based on our hydrodynamical simulation compared to the observed photometry is shown in Fig.~\ref{fig:SED}. The stellar spectrum is taken from the Castelli--Kurucz atlas \citep{castelli2004}, choosing the model for a stellar type of G8V (T$_{\mathrm{eff}}=5500$\,K), and a surface gravity of $\mathrm{log}\,g=4.5$. The flux measurements from the VLT Interferometer (VLTI)/PIONIER, SPHERE/IRDIS/IFS, 2MASS, IRAS, WISE, and AKARI hint at a strong NIR and mid-IR (MIR) excess expected for a disk around a young star. Although our model lies within the measurement uncertainties of the SPHERE data points, there is a clear trend that it underestimates the fluxes between $\sim$5 and $\sim$50\,$\mu$m. The explanation is that our model only considers the dust disk from our hydrodynamical simulations with a radial grid starting at $R_{\mathrm{in}}=18$\,au. The inner disk region is hidden behind the coronagraph for our SPHERE data, so there is no strong constraint on the geometry and orientation of an inner dust belt or halo. This makes the modelling of the inner disk regions very degenerate, and thus, in our model there is no dusty material close to the star. In general, the inner radius regulates the maximum temperature of the dust grains at the inner rim. For larger $R_{\mathrm{in}}$ the luminosity excess shifts from the NIR to the MIR \citep[see e.g.][]{woitke2016}. Exploring a larger grid of dust and disk structure parameters is beyond the scope of the paper as we did not attempt to do a detailed fit of the SED. The photometric point in the mm regime is nicely reproduced with our simulated flux. Due to the absence of additional photometric points in the (sub-)mm, the mm slope fit cannot be evaluated. For example, the dust size distribution power-law index (typical value of 3.5 in our case for both populations) changes, in particular, the mm and cm slopes. Although our current model indicates a good disk mass estimate, this situation might change when additional dust is included in the inner disk regions in order to fit the NIR/MIR fluxes. The underestimate of this emission could result in a slight error in dust temperature and requires a higher disk mass. This is certainly compatible with the disk mass estimate from \citet{ansdell2016} as this value only gives a lower limit.

\section{Conclusions} \label{sec:conclusions}
We observed the large circumstellar disk around RY~Lup with SPHERE/IRDIS in scattered light with high angular resolution using polarimetric and angular differential imaging and uncovered directly for the first time an inclined disk with spirals. The disk position angle is 107 deg and its inclination is 70 deg. Our observations show the complementarity between polarimetric and angular differential imaging. Polarimetric observations allow the study of the disk without the self-subtraction effects inherent to ADI-processed data, while the ADI observations provide deeper detection limits further away from the star, enabling us to study the disk and search for exoplanets. We also retrieved a high signal-to-noise ratio spectrum for the disk using the IFS. The disk colour is grey, which is an indication that dust grains larger than the wavelength dominate the scattering opacity in the disk surface.

We have studied the morphology and surface brightness of the disk and formulated a hypothesis on the origin of the spiral arms. An explanation for the spiral arms could not be uniquely determined due to the high inclination of the disk. Our numerical model shows that the spiral arms could be explained by one low-mass planet (2 M$_\mathrm{jup}$) located at 190 au orbiting exterior to the spiral arms. Although this model describes the formation of spiral arms qualitatively similar to the features in the NIR scattered light observations of RY~Lup, it is only a suggested configuration for the system, and not a best-fitting model. However, given the high inclination of the disk it is unlikely that the planet's thermal radiation is directly detectable. While spirals can be excited by the tidal interaction with the companion, they can also be triggered by the close proximity of shadows in the disk as discussed in \citet{montesinos2016,benisty2017}. Also, we cannot exclude that the 0.8$\arcsec$ diameter gap detected in sub-mm continuum emission could host additional massive planets that could play a role in the disk morphology. 
The innermost gap discovered in the ALMA observation cannot be detected in scattered light due to the high inclination of the disk. The scattered light flux shows a small profile asymmetry which does not coincide with the symmetry of the sub-mm continuum emission. This is likely the result of surface density perturbation related to the presence of the spiral arms and could be strongly related to the viewing orientation.

Our observations are compatible with the hypothesis made by \cite{manset2009} of photometric and polarimetric variations created by an almost edge-on circumstellar disk that is warped (or inclined) close to the star where it interacts with the star magnetosphere. This configuration could lead to a partial shadowing of the outer disk and a brightness asymmetry that can remain undetected due to the high inclination of the disk. We also performed a detailed radiative transfer study, which reproduce well our scattered light observations. The planet--disk interaction scenario is in agreement with the ALMA high-resolution  dust continuum image being truncated at $\sim$120\,au. This modelling effort cannot fully constrain all the parameters in this case, but helps to understand the disk structure and distribution of small grains in particular. 

\begin{acknowledgements}
We acknowledge our anonymous reviewers for the very careful reading and constructive suggestions that contributed to improving this paper.
SPHERE is an instrument designed and built by a consortium consisting of IPAG (Grenoble, France), MPIA (Heidelberg, Germany), LAM (Marseille, France), LESIA (Paris, France), Laboratoire Lagrange (Nice, France), INAF - Osservatorio di Padova (Italy), Observatoire astronomique de l'universite de Geneve (Switzerland), ETH Zurich (Switzerland), NOVA (Netherlands), ONERA (France), and ASTRON (Netherlands) in collaboration with ESO. SPHERE was funded by ESO, with additional contributions from CNRS (France), MPIA (Germany), INAF (Italy), FINES (Switzerland), and NOVA (Netherlands). SPHERE also received funding from the European Commission Sixth and Seventh Framework Programmes as part of the Optical Infrared Coordination Network for Astronomy (OPTICON) under grant number RII3-Ct-2004-001566 for FP6 (2004-2008), grant number 226604 for FP7 (2009-2012), and grant number 312430 for FP7 (2013-2016). This work was supported by the Programme National de Plan\'etologie (PNP) and the Programme National de Physique Stellaire (PNPS) of CNRS-INSU co-funded by CNES. This work has also been supported by a grant from the French Labex OSUG2020 (Investissements d'avenir - ANR10 LABX56) and by the ANR grant ANR-14-CE33-0018 (GIPSE). This work has made use of the SPHERE Data Centre, jointly operated by OSUG/IPAG (Grenoble), PYTHEAS /LAM/CeSAM (Marseille), OCA/Lagrange (Nice), and Observatoire de Paris/LESIA (Paris). This research has made use of the VizieR catalogue access tool, CDS, Strasbourg, France. Q. Kral also acknowledges support from STFC (consolidated grant) via the institute of Asronomy, Cambridge.
\end{acknowledgements}

\bibliographystyle{aa}
\bibliography{rylup}
\end{document}